\documentclass[useAMS,usenatbib]{mn2e}
\pdfoutput=1

\usepackage{amssymb}
\usepackage{amsmath}
\usepackage{graphicx}
\usepackage{xspace}
\usepackage{float}
\usepackage[draft]{hyperref}
\usepackage[usenames,dvipsnames,svgnames,table]{xcolor}

\newcommand{\Msol}{\mathrm{M_{\sun}}}
\newcommand{\LCDM}{$\Lambda$CDM\xspace}

\newcommand{\galform}{\textsc{galform}\xspace}
\newcommand{\subfind}{\textsc{subfind}\xspace}
\newcommand{\Nbody}{$N$-body\xspace}

\title[Making Mocks of the Stellar Halo]{Creating mock catalogues of stellar haloes from cosmological simulations} 
\author[Ben Lowing et al]{Ben Lowing$^{1}$\thanks{E-mail:
    b.j.lowing@durham.ac.uk}, Wenting Wang$^{1}$\thanks{E-mail:bilinxing.wenting@gmail.com},  Andrew Cooper$^{1,2}$, Rachel
  Kennedy$^{1}$, John Helly$^{1}$,  \newauthor Shaun Cole$^{1}$ and Carlos
  Frenk$^{1}$ \\
$^{1}$Institute for Computational Cosmology, Department of Physics, University of Durham, South Road, Durham, DH1 3LE, UK \\
$^{2}$National Astronomical Observatories, Chinese Academy of Sciences, 20A Datun Road, Chaoyang, Beijing 10012, P.R. China}

\begin{document}

\date{Accepted ... Received ... in original form ...}
\pagerange{\pageref{firstpage}--\pageref{lastpage}} \pubyear{2013}
\maketitle


\label{firstpage}

\begin{abstract}
We present a new technique for creating mock catalogues of the
individual stars that make up the accreted component of stellar haloes
in cosmological simulations and show how the catalogues can be used to
test and interpret observational data. The catalogues are constructed
from a combination of methods. A semi-analytic galaxy formation model
is used to calculate the star formation history in haloes in an \Nbody
simulation and dark matter particles are tagged with this stellar
mass. The tags are converted into individual stars using a stellar
population synthesis model to obtain the number density and 
evolutionary stage of the stars, together with a phase-space sampling
method that distributes the stars while ensuring that the phase-space
structure of the original \Nbody simulation is maintained. A set of
catalogues based on the \LCDM Aquarius simulations of Milky Way 
mass haloes have been created and made publicly available on a website. 
Two example applications are discussed that demonstrate the power and 
flexibility of the mock catalogues. We show how the rich stellar 
substructure that survives in the stellar halo precludes a simple 
measurement of its density profile and demonstrate explicitly how 
pencil-beam surveys can return almost any value for the slope of the 
profile. We also show that localized variations in the abundance of 
particular types of stars, a signature of differences in the composition 
of stellar populations, allow streams to be easily identified.

\end{abstract}


\section{Introduction}

Extensive diffuse stellar haloes are known to surround both the Milky
Way and M31 \citep[e.g.][]{Deason:2011,Ibata:2014}. The $\Lambda$CDM model of
hierarchical structure formation predicts that these haloes should be a
ubiquitous outcome of galaxy formation resulting from the accretion of
satellite galaxies that fall into the potential of the larger host galaxy and
are subsequently disrupted \citep{Bullock:2005, Cooper:2010}. The dynamical
time of the stellar halo is very long compared to the age of the host galaxy
and so it preserves a memory of its initial conditions. From the
identification of distinct populations within haloes, and the study of
properties such as kinematics or chemical composition, a vast amount of
information about the assembly history of the galaxy can be inferred. 

The outer regions of stellar haloes are dominated by multiple, extensive tidal
streams, while the inner regions, where dynamical timescales are shorter,
experience a more complete destruction of substructure and therefore have a
smoother stellar distribution. In addition to the accreted component, stellar
haloes are thought to contain a second component made up of stars that formed
{\em in situ}. Recent observational claims of a two-component halo in
the Milky Way have focused attention on this possibility 
\citep[e.g.][]{Carollo:2010, Beers:2012,Kafle:2013}, while the evidence 
for two components is disputed by \citealt{Schoenrich:2014}. From a theoretical 
perspective there is a wide variety of possible formation mechanisms for 
{\em in situ} halo stars, many of which overlap with possible origins of the Galactic 
thick disc.  These include the scattering of stars from the thin disc, the collapse
of the disc at high redshift, star formation in gaseous tidal streams and
thermal instabilities in a hydrostatic gas halo \citep[e.g.][Parry et al. in
prep.]{Abadi:2006,Zolotov:2010,Font:2011a,Tissera:2013}.  Most of these
mechanisms result in {\em in situ} components that dominate in the inner few
kiloparsecs of the galaxy and remain important out to $\sim30$kpc. 

Over the last few years there has been much effort dedicated to modelling
stellar haloes \citep[e.g.][]{Bullock:2005, Cooper:2010, Zolotov:2010,
Font:2011a, Tissera:2012} motivated by improvement in observations. Recent
surveys such as {\sc sdss segue} \citep{Yanny:2009} and {\sc p}an-{\sc starrs}
\citep{Kaiser10} have started to probe deeper into the Milky Way halo and are
starting to build up a picture of its complex structure. The {\sc
gaia} mission will lead to further major improvements in the Milky Way data
\citep{Jordi:2010, de-Bruijne:2012} and new photometric surveys are extending
these studies to other nearby galaxies \citep{McConnachie:2009,
MartinezDelgado:2010}. We are now at the point where careful comparisons
between the haloes predicted by theory and those observed in the universe are
becoming possible.

In order to make a direct comparison of theoretical predictions with
observations of stellar haloes, a model of stellar halo formation is required,
along with a way to convert the output into observable quantities. Furthermore,
to obtain realistic accretion histories, a complete model of galaxy formation
in a cosmological context is essential. Various studies have focused
on supplementing \Nbody simulations with analytical modelling of star formation
and chemical enrichment
\citep[e.g.][]{Bullock:2001,Robertson:2005,Font:2006,Johnston:2008,DeLucia:2008,Tumlinson:2010}.
Alternatively, cosmological hydrodynamical simulations have also been used to
study the overall structure and assembly of stellar haloes
\citep[e.g.][]{Crain09,Zolotov:2010,Libeskind:2011,Tissera:2012,Pillepich:2014},
although these do not yet have the capability to resolve the detailed kinematic
and spatial structure of individual halo stellar populations. Even ``zoom''
simulations of individual galaxies represent the stellar halo with a relatively
small number of star particles, since this component amounts to only a few per
cent of the total stellar mass.

In this paper we employ the particle tagging method developed by
\citet[hereafter C10]{Cooper:2010} to construct stellar halo
models. The method is based on high-resolution dark matter simulations
in which the star formation occurring in haloes is calculated using a
semi-analytic galaxy formation model. At every output time in the
simulation, an appropriate set of dark matter particles is tagged with
the stellar mass that formed since the previous output time. The
advantages of this technique are that it can create models of much
higher resolution and at a much lower computational cost than full
hydrodynamical simulations. C10 demonstrated that the method produces
complex stellar haloes containing tidal streams, shells and other
substructure, whose ages and metallicities are in broad agreement with
recent observations.

A limitation of the C10 tagging method is that ``star particles'' are not
individual stars but rather represent entire stellar populations which can
contain thousands of solar masses of stars. The challenge is to compare these
sparsely sampled haloes to actual observational surveys which often focus on a
particular type of star, such as blue horizontal branch (BHB) stars. These may
not trace the overall stellar mass distribution in an unbiased way. To overcome
this limitation, we use a method similar to that employed in {\sc galaxia}
\citep{Sharma:2011} to construct customizable mock catalogues of halo stars.
Each star particle is treated as a separate simple stellar population using
theoretical isochrones to determine the abundance of each type of star expected
in a population of the corresponding age and metallicity. To create positions
and velocities of individual stars, the star particle distribution is
oversampled using a phase-space kernel method based on the EnBiD code
\citep{Sharma:2006} which maintains the underlying phase-space structure.

Our mock catalogues allow us to compare the structure of simulated
stellar haloes to observational data in a more realistic way. For example,
obscuration by the Milky Way disc, survey limits imposed by the high
cost of deep photometry and spectroscopy and the low surface density of halo
tracers mean that the structure of the Galactic stellar halo is only well
sampled to large distances in small patches of the sky. Nevertheless, attempts
are often made to infer the profile of the Milky Way stellar halo from these 
limited-area observations \citep[e.g.][]{Watkins:2009,Sesar:2011,Akhter:2012}. 
In this paper we show how the complicated uneven structure of the outer accreted
stellar halo results in large fluctuations among measurements of its
global properties based on observations of small sky patches. We also show how
different satellites that build up the halo create observable variations in the
stellar populations across the sky, for example in the ratio of stellar types.
This variation can potentially help to detect new halo substructures in the
Milky Way.

In Section~2 we outline the methods used to create stellar halo
models. The final output of the process, a set of mock catalogues of
individual stars and their properties, is described in further detail
in Section~3. Sections~4 and~5 present two simple example applications
of the mock catalogues to study the overall structure of stellar
haloes in our simulations and the distribution of different stellar
types over the sky. Finally, in Section~6 we discuss some potential
advanced applications of the mock catalogues and how they can be used
to interpret and test observational strategies.

\section{Method}

Creating a model of stellar haloes based on a consistent theory of galaxy
formation in a cosmological context requires a combination of various
techniques. The foundation of our models is a set of \Nbody dark matter only
``zoom'' cosmological simulations of Milky Way mass haloes onto which we 
have grafted the Durham semi-analytic model of galaxy formation, \galform \citep{Cole00}.
\galform \ predicts the amount of stars that form in every halo and subhalo at
each output time, along with an estimate of the metallicity of each stellar
population. This stellar mass is then assigned to dark particles within the
\Nbody simulations using the C10 particle tagging technique, which allows us to
track the accretion and disruption of satellites and follow the fate of the
stripped stars. Finally, to convert the massive simulation particles
into mock catalogues (mocks for short) of individual stars, we use a
method based on stellar population modelling and phase-space sampling. Each of
these steps is briefly outlined in the following subsections.

\subsection{\Nbody Simulations}

For this work we use the haloes from the Aquarius project, which simulates six
dark matter haloes of mass $\sim 10^{12}\,$M$_{\odot}$ at multiple resolution
levels \citep{Springel:2008a, Springel:2008b, Navarro:2010}.  The simulations
assume the standard \LCDM cosmology with parameters chosen to be consistent
with the results from the {\em WMAP} 1-year data \citep{Spergel:2003} and the
2dF Galaxy Redshift Survey data \citep{Colless:2001}: matter density parameter,
$\Omega_{\rm M} = 0.25$; cosmological constant, $\Omega_{\Lambda} = 0.75$;
power spectrum normalization, $\sigma_8 = 0.9$; spectral index, $n_{\rm s} =
1$; and Hubble parameter $h=0.73$. The six haloes were selected randomly from a
set of isolated $\sim 10^{12}\,$M$_{\odot}$ haloes identified in a lower
resolution ($900^3$-particle) parent simulation of a $100 h^{-1}$ Mpc cubic
volume \citep{Gao:2008}. The isolation criterion requires a halo to have no
neighbours exceeding half its mass within $1 h^{-1}$~Mpc. This weak
selection criterion ensures that the haloes are not members of any massive
groups or clusters.

The Aquarius simulations are labeled Aq-A to Aq-E; in each case we use
the second highest resolution (``level-2'' in the Aquarius notation)
simulation, with a particle mass of at most $10^4 h^{-1}\,$M$_{\odot}$. We have
used only five of the six Aquarius haloes, omitting halo Aq-F which undergoes
two major mergers at $z \sim 0.6$ and is thus highly unlikely to host a disc
galaxy at $z=0$. For each simulation we have 128 output times, with a constant
spacing of 155~Myr after $z\approx2.5$.

\subsection{Galaxy Formation Model}

Dark matter $N$-body simulations are ideal for following the non-linear growth
of cosmic structure, including the detailed accretion history of a halo.
However, including a separate, realistic treatment of baryons in a
cosmological simulation at sufficient resolution for stellar halo studies is
still computationally prohibitive. Semi-analytic galaxy formation modelling
permits baryonic physics to be included in much higher resolution $N$-body
simulations at low computational cost, and is thus ideal for this purpose.
Based on simple theoretical treatments and empirical prescriptions,
semi-analytic models can provide a good match to a remarkable number of
observations of the local and distant galaxy populations \citep{Baugh:2006}.
These models, however, are not designed to follow the internal structure of
galaxies in detail, hence the need for the particle tagging extension described
below.

The Durham semi-analytic model, \galform, is used in this work
to postprocess the Aquarius \Nbody simulations. \galform computes the
mass of stars that forms in each halo between every two successive simulation
outputs, along with the properties of these stars, such as their metallicity.
Rather than using the \citet{Bower:2006} model on which C10 was based, we use
the more recent \citet{Font:2011b} model, which includes a modified treatment
of physics relevant to satellites. The only significant changes compared to the
C10 model are the use of a higher yield, a modified feedback model for
supernovae in which the mass ejection efficiency saturates in haloes with
circular velocity $V_{\rm circ} \leq 65$ km~s$^{-1}$, and an earlier epoch of
hydrogen reionization. The \citet{Font:2011b} model produces a satellite
luminosity function that matches that of Milky Way and also generates stellar
populations in satellites which match the observed luminosity-metallicity
relation. This eliminates a shortcoming of C10 where the overall metallicity of
the stellar halo was found to be significantly lower than that of the Milky
Way's stellar halo.

\subsection{Particle Tagging}
\label{sec:tagging}

The C10 particle tagging method is a means to associate the stellar
mass calculated by \galform with six-dimensional phase space volumes, defined
by carefully chosen sets of representative particles in high-resolution dark
matter only simulations. This allows the 3D spatial distribution and kinematics
of stellar populations to be followed, albeit approximately, without including
gas physics in the original simulation.

After applying \galform to the outputs of the Aquarius simulations,
C10 associated each newly formed stellar population (discretized in age by
simulation output times) with the 1\% most tightly bound dark matter particles
in their corresponding dark matter haloes. These sets of `tagged' dark matter
particles are chosen at the time when the population forms. Their diffusion in
configuration and velocity space can then be tracked to the present day. This
method is ideal to study the formation of the outer accreted component of
stellar haloes, which are thought to be created by the tidal disruption of
satellites. The fraction of most-bound particles tagged, the free parameter
introduced by the method, has an effect on the scale radii of the resulting
galaxies. C10 fixed this value at 1 per cent in order to reproduce the
distribution of observed half-mass radii for surviving satellites. Having fixed
this scale, satellite profiles and velocity dispersions also match well to
observations (see C10).

Most surviving satellites are highly dark matter dominated
\citep[e.g.][]{Walker:2007}. Since the $N$-body potential is not altered by the
gravity of the `stars' represented by the tags, the technique is likely to be
much less accurate in baryon-dominated systems, such as the centre of the Milky
Way analogue (see C10, \citealt{Cooper:2013} and \citealt{Cooper:2014} for
further details and discussion). Satellite growth rates, survival times and
orbits may also be affected by a more self-consistent treatment baryons
\citep[e.g.][]{Sawala13}. The relative importance of these effects is
uncertain, however. For example, supernova feedback can create dark matter
cores in haloes \citep[e.g.][]{Navarro:1996,Pontzen:2013}. This 
reduction in central density may reduce their survival time when they are 
subjected to tidal stripping \citep[e.g.][]{Penarrubia:2010}. On the
other hand, the central concentration of stars may also bind these galaxies
more tightly in a self-consistent simulation -- in this context we note that
numerical softening of the gravitational force between particles creates
artificial cores in our simulated haloes on scales of $\sim100$~pc.

To explore the consequences of such limitations, \citet{Bailin:2014}
compared tagged dark matter particles in hydrodynamical simulations to star
particles formed self-consistently in the same simulations. They found moderate
disagreement in the distributions of halo stars represented by these two sets
of particles, and concluded that tightly bound dark matter is not a reliable
proxy for the dynamics of stars in satellite galaxies. However, there is a
fundamental difference between the tagging method tested by \citet{Bailin:2014}
and that used by C10.  \citet{Bailin:2014} carried out a single tagging
procedure for each satellite at the time of its infall into the main halo,
whereas C10 tag each stellar population at the time of its formation.  Star
particles undergo considerable diffusion in energy over time, particularly if
their host halo is disturbed, and hence would not be expected to occupy the
same region of phase space as tightly bound dark matter chosen long after they
form.  Dark matter particles with similar binding energy chosen at the time of
star formation, on the other hand, will undergo similar diffusion to star
particles.  This diffusion is therefore captured in the C10 approach, but not
in the method used by \citet{Bailin:2014}. Two forthcoming papers will test the
tagging technique of C10 in a similar way (Le Bret et al. and Cooper et al.  in
prep.). Within the limitations and approximations of the method, mentioned
above and in C10, these new studies find good correspondence between stars and
tagged dark matter particles in hydrodynamical simulations.

\subsection{Mock Catalogue Construction}

The final step is to turn the output of the tagged simulations into full mock
catalogues of individual stars. The output of the model is a collection of
$N$-body particles, each tagged with some amount of stellar mass, and having
both an age for when the stars were born and a metallicity at that time. When a
single dark matter particle is tagged multiple times, multiple `tags'
are created with the same phase space trajectory. Turning these tags into a
full mock catalogue requires two steps. The first is to determine the total
number of stars of different types that each tagged particle represents by
considering each as a single-age stellar population, and the second step is to
distribute the stars over the phase-space volume represented by the
corresponding dark matter particle. The following subsections discuss each
step in detail.

\subsubsection{Stellar Population Synthesis}

Unfortunately, even with the high-resolution Aquarius simulations we are a long
way from being able to represent each individual star by a single $N$-body
particle. Instead, in the C10 simulations, each stellar particle can represent
up to $\sim10^{3} \mathrm{M_{\sun}}$. In order to convert these massive stellar
particles into individual stars we consider each tagged particle as a stellar
population (SSP), in which all the stars formed at the same time from gas with
the same metallicity; the theoretical justification of this is discussed in
detail in \citet{Pasetto:2012}. The abundances of the different star types that
make up the population can then be obtained using stellar population synthesis
modelling.

The initial mass function (IMF) determines the number of stars forming within
each stellar population as a function of initial stellar mass. Here we adopt a
Kennicutt IMF with no correction for brown dwarfs \citep{Kennicutt:1983}, just
as was used in the semi-analytical model, to ensure consistency. Theoretical
stellar isochrones then specify whether each type of star still survives at the
final time, along with the current properties of these stars.

Stellar isochrones allow us to assign properties such as temperature, magnitude
and colour to stars of a given initial mass, age and metallicity. We employ the
{\sc parsec} isochrones \citep{Bressan:2012}, as these are recent, up-to-date
isochrones that extend over a wide range of metallicities $(0.0001\leq Z \leq
0.06)$. We create a grid of isochrones spanning the range of ages between $6.63
< \log_{10} t/{\rm yr} < 10.13$, with a step size ($\Delta \log(t)=0.0125$),
and a range of metallicities between $0.0001\leq Z \leq 0.005$. Interpolating
between ages and metallicities would require matching different evolutionary
points along the isochrones, so instead we simply identify the nearest
isochrone in the grid and associate that with a tag. A finer grid could be used
if required, but the adopted resolution was found to be adequate for our
current purposes.

The method can be summarized as follows: for each tag, split the population
into a range of initial stellar mass bins. Use the IMF to determine the
fractional weighting of each bin. Then find the closest isochrone in age and
metallicity and use that to assign properties to the stars in each bin.
Finally, multiply the mass of the tag by the weight of the bin to
obtain the total mass and, from that, the number of stars in that bin. Since
the lifetimes of massive stars are much shorter than the age of the Universe,
only a fraction of the initial stellar mass of the tag will be converted to
stars in the mock catalogue, the balance being returned to the interstellar
medium or locked in stellar remnants.
	
\subsubsection{Phase space Sampling}

\begin{figure*}

\centerline{\includegraphics[width=1.0\textwidth]{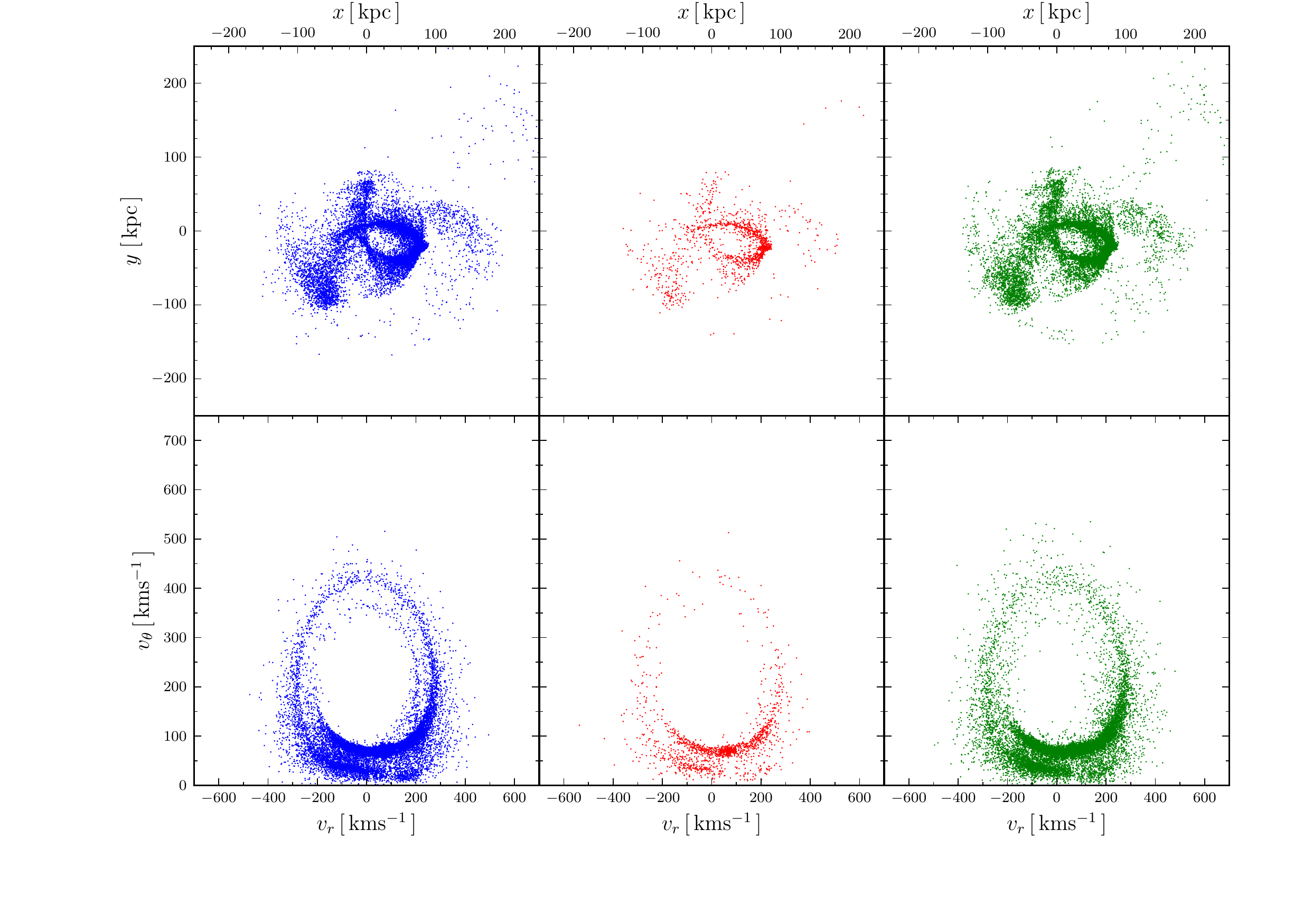}}

\caption{Real-space projection of an example satellite (upper row) and
velocity-space projection (bottom row). The left column shows the original set
of satellite particles, while middle column  is the down-sampled version with
just 10 percent of the number of particles. The right column shows the
result of resampling 10 per cent of the particles back to the original
resolution using phase space kernels based on EnBiD density estimates.}

\label{fig:satelliteProjections} \end{figure*}

The next step is to split the massive stellar particles representing SSPs into
stars with individual positions and velocities. Although the massive stellar
particles provide a coarse sampling of the phase-space distribution of the
stellar halo, it is desirable to smooth the distribution and increase the
sampling. This is done by distributing stars over the entire phase-space volume
that each tagged particle represents. Using the phase-space volume rather than
just the real-space volume preserves coherent structures such as tidal streams.
It must be noted that increasing the sampling does not increase the resolution
and cannot reveal any more detailed structure than was in the original
simulation.

To estimate the phase-space volume each stellar mass particle occupies
we have used the entropy-based binary decomposition (EnBiD) code of
\cite{Sharma:2006}. EnBiD numerically estimates the densities of
discretely sampled data based on a binary space partitioning
tree. EnBiD uses an entropy-based node splitting criterion to decide
which subspace (real or velocity space) to split next. This makes the
method metric free, as there is no need to specify in advance how
velocity and configuration spaces relate to each other, and ensures
the subspace with greater variation is subdivided more.

Once the binary tree has been built, the leaf nodes can be used to
estimate the phase-space volume of each particle. The volume of the
hypercube of the leaf nodes can vary significantly even between close
neighbours due to Poisson sampling noise of the original particle
distribution. Simply using the volumes of raw nodes would result in a
large dispersion, but this can be reduced by using a smoothing
scheme. We have modified EnBiD's anisotropic kernel density estimation
scheme to return a volume as well as a density. The scheme works by
first calculating a local metric about each particle based on the
shape of the hypercube of each leaf node and using the length of the
six sides to scale the phase space in each direction. A covariance matrix
is then calculated based on the nearest 64 neighbours in the scaled
space. Diagonalizing the covariance matrix gives a set of eigenvalues
and eigenvectors that can be used to perform a further transformation
to a coordinate system in which the eigenvectors define the principal
axes. In this new rotated and scaled space the local particle
distribution will have unit covariance in each direction. Finally, the
phase-space volume of the particle is defined as a hypersphere with
1/40 of the volume of a hypersphere based on the distance to the 40th
nearest neighbour, $R_{40}$. It therefore has a radius $R_1$ of
\begin{equation}
R_1 = \left( \frac{1}{40} \right)^{1/6} R_{40}.
\end{equation}
The modified version of EnBiD returns a 6x6 matrix that describes the
phase-space volume of each particle.

The actual sampling is performed by choosing points from an 6d isotropic
Gaussian distribution in which each component has zero mean and variance
$\sigma^2=\gamma R_1^2$. Each generated point is then transformed by applying
the inverse operations to convert from the sampled space back to positions and
velocities in the original simulation. Some care has to be taken with the
choice of the scale parameter $\gamma$ which relates the scale of the Gaussian
kernel to the scale of the hypersphere returned by EnBiD. Choosing too large a
value of $\gamma$ will result in the phase space structure of each component of
the halo being made hotter and too diffuse. Choosing too small a value of
$\gamma$ will result in an artificially clumpy phase space distribution. As a
compromise we have chosen $\gamma=1/48$.  In a separate study (Wang et al in
preparation) we have developed a maximum likelihood method for modelling the
distribution function of the stellar halo in order to estimate the mass of the
host dark matter halo. Using this approach we have found that the recovered
halo mass is biased high if $\gamma$ is too large. With $\gamma=1/48$ the 
level of the bias is controlled
to be less than about $10$\%, which is much smaller than typical measurement
errors. Smaller values of $\gamma$ would remove this bias entirely, but would
negate the benefit of distributing stars over a phase space kernel,
because in the limit $\gamma \rightarrow 0$ one is merely placing all the stars
at the same phase-space location as the tagged particle from which they arise.

In order to test our EnBiD sampling method and demonstrate that we have made a
suitable choice for the parameter $\gamma$, we attempt to reconstruct a
debris stream from one of our simulations after artificially degrading its
resolution by sampling only a fraction of its particles. We extracted from the
simulation a satellite in the process of undergoing significant tidal
stripping, such that its particles have an extended distribution in both
configuration and velocity space. This object comprises 15,250 particles. A
lower resolution version was created by randomly selecting 10 per cent of the
particles. Finally, the low resolution version was then resampled using our
EnBiD method, with each particle being converted back into 10 particles, thus
creating an object with the same resolution as the original satellite. A
comparison between the two then demonstrates the success of our phase space
sampling method.

Fig.~\ref{fig:satelliteProjections} shows projections of phase space for the
different versions of the satellite. The top row is a real-space projection and
the bottom row a velocity projection. The original satellite (left column) has
a large amount of structure in both real and velocity space, with several
complete wraps of the tidal tails visible. In the down-sampled version (the
second column from the left) the primary features are still present but the
subtle ones are missing, while the remnant of the satellite core is more
prominent.  The EnBiD up-sampled version (the third column from the left) is
remarkably similar to the original satellite, though the subtle features tend
to be smoothed out. In particular the arc in velocity space
corresponding to an orbital apocentre at $v_{\theta} \approx
400\,$km~s$^{-1}$, while still present, is less well resolved.  This is
expected, as information is lost during down-sampling and no resampling scheme
will be able to recover it perfectly.

\begin{figure}
\centerline{\includegraphics[width=1.0\linewidth,clip=True,trim=0cm 1cm 0cm 0.5cm]{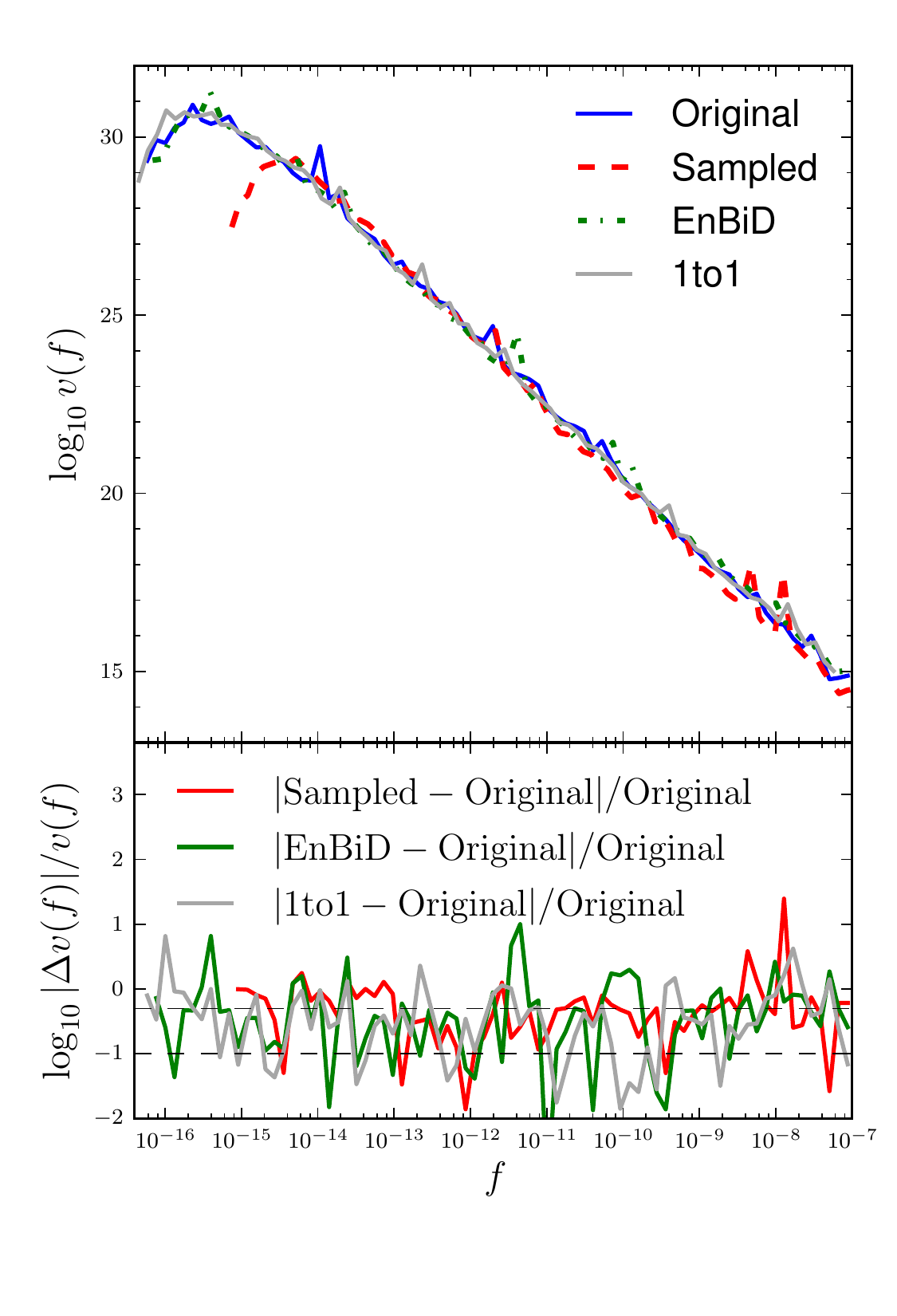}}

\caption{Top: Comparison of the volume distribution function of
phase-space density, $v(f)$, for different phase space sampling schemes applied
to a single satellite. Both axes have arbitrary units. Blue (labeled
`original'): the original high-resolution particle distribution. Red (`sampled'): 
a sample of $1/10$ of the original distribution. Green (`EnBiD'): the same number 
of particles as the original distribution, drawn from the phase space kernels 
defined by the down-sampled (red) distribution. Grey(`1-to-1'): each particle resampled once from 
the phase space kernels defined by the original distribution. Bottom: Relative difference 
between the volume distribution functions of the `sampled', `EnBiD' and `1-to-1' 
results to the `original' volume distribution function. Black dashed lines mark the 10\% and 50\% 
level of difference.}

\label{fig:enbidPhasespaceVolDist}
\end{figure}

A further test of how well the original and EnBiD versions match is a
comparison between the volume distribution function of phase-space density,
$v(f)$, as shown in the upper panel of
Fig.~\ref{fig:enbidPhasespaceVolDist}. We have used the technique of
\citet{Arad:2004}, based on Delaunay tessellation, as an independent
measurement of the phase-space density. The phase space volume distribution
function is found to be similar in all three cases, suggesting that it is
robustly defined by even a small subset of the particles and that the EnBiD
resampling process does not significantly alter it. The differences
can be seen more clearly in the bottom panel of
Fig.~\ref{fig:enbidPhasespaceVolDist}.  The relative difference between the
down-sampled volume distribution function and the original is shown by the red
curve, while the relative difference between the EnBiD up-sampled version and
the original is show by the green curve. We have also tested how much noise is
introduced to $v(f)$  simply by randomly sampling the phase space distribution
defined by the original particles (as estimated by EnBiD) at a ratio of $1:1$
(grey line). The relative difference in all these cases is mostly within 10 to
50 per cent, as indicated by the two horizontal black dashed lines. There is no
systematic trend in the relative difference across 9 orders of magnitude in
phase space density. A few spikes correspond to larger differences in
particular regions, which appear to be almost entirely due to the shot noise
introduced by the discreteness of the particle distribution.

We also notice that at the low density end of
Fig.~\ref{fig:enbidPhasespaceVolDist} the red dashed curve based on the
down-sampled particles starts to plunge at $f \sim 10^{-14.5}$, whereas the
blue curve based on the original sample continues to increase down to $f \sim
10^{-15.5}$. The down-sampled particles are more sparsely distributed in phase
space and thus sample low density regions poorly, as well as increasing the
Poisson noise.  The green dot-dash curve, which is based on the particles
resampled with EnBiD, recovers the same small scale behaviour of the original
particle distribution. The grey curve, where each particle is resampled once 
from the phase space kernel of the original sample, have almost the same small scale 
behaviour of the original sample as well.

When applying this resampling technique to the tagged particles an important
consideration is how the particles are divided into subsets on which to run
EnBiD. The particular subset that is chosen defines which particles are
considered as phase-space neighbours. We choose to apply the EnBiD
volume estimation to all the tagged particles that form along
a given branch of the halo merger tree. This separates the tagged particles
into sets based upon the halo they formed in.

It is also possible to split the tagged particles by individual
assignments, i.e. only particles from an individual single-age stellar
population are considered as potential phase-space neighbours. This has the
advantage that the phase-space volumes of particular populations are not
diluted by stars forming in the same halo at later times. The problem with this
method is that sets of particles from individual assignments can be very small
in many cases, such that their phase-space volumes are too poorly sampled to
accurately constrain the structure in six dimensions. While our implementation
allows either method to be used, we have chosen to use the first method in the
rest of this paper.

EnBiD requires a minimum of 64 particles for the covariance matrix
calculation. We therefore exclude assignments with smaller numbers of
particles than this. We have checked that this eliminates a negligible
amount of stellar mass.

\section{Mock catalogues}

In this section we describe the range of information provided by our mock
catalogues, and present a set of example mocks generated from the Aquarius
simulations. The combination of the various techniques used results in a rich
dataset. 

\subsection{Public Catalogues}
\label{sec:cat}
\begin{table*}
\center
\begin{tabular}{l c l l}\hline \hline
Field & Data type & Unit & Description \\ \hline
ID & long & & ID of the star, unique within the mock catalogue \\
Position $(x, y, z)$ & float & kpc & Star position, relative to halo centre \\
Velocity $(v_x, v_y, v_z)$ & float & km~s$^{-1}$ & Stars velocity, relative to halo rest frame \\
Galactic coordinates $(l, b)$ & float & $^o$ & Galactic longitude and latitude relative to solar observer \\
Radial distance $(r)$ & float & kpc & Distance from solar observer \\
Relative velocity $(v_r, v_{\theta})$ & float & km~s$^{-1}$ & Radial and tangential velocity relative to solar observer \\
$\log_{10}\mathrm{(Age)}$            & float & $\log_{10}$(yr)               & Age of star \\
Metallicity    & float &                   & Fraction of star mass in metals \\
Mass           & float & $\Msol$           & Current mass of star \\
$\log_{10}(g)$ & float & $\log({\rm cgs})$ & Surface gravity of the star \\
$\log_{10}(T_{\mathrm{eff}})$ & float & $\log_{10}(\mathrm{K})$ & Effective temperature of the star \\
Type & int &  &  The original PARSEC evolutionary stage flag that labels the type of star  \\
& & & (0=Pre-MS, 1=Main sequence, 2=Subgiant branch, 3=Red giant branch,   \\
& & & 4,5,6=stages of core He burning, 7,8=stages of Asymptotic Giant Branch)\\
$M_{(u,g,r,i,z)}$ & float & & Absolute rest frame ($u$,$g$,$r$,$i$,$z$) band
({\sc sdss}) magnitude of the star \\
$m_{(u,g,r,i,z)}$ & float & & Apparent rest frame ($u$,$g$,$r$,$i$,$z$) band
({\sc sdss}) magnitude of the star \\
& & & as seen by the solar observer \\
Dark Matter ID & long & & The tagged dark matter particle this star was spawned from \\
Subhalo ID & long &        & The current subhalo to which the associated dark matter particle \\
& &                                      &  belongs to. -1 if not currently a member of any subhalo, 0 if part \\
& &                                      & of the main halo \\
Tree ID & long &              & The subhalo tree that this star came from. All stars that  arrive in the\\
& &                                     & main halo via the same satellite have the same ID \\
Infall Redshift & float & & Accretion redshift of the subhalo this star once belonged to. Defined as the \\
& & & redshift at which the satellite reaches its maximum mass (see text). \\
\hline
\end{tabular}
\caption{Information provided in our online mock catalogues that may 
  be accessed at \url{http://galaxy-catalogue.dur.ac.uk:8080/StellarHalo}. 
SQL can be used to query any of these fields.
\label{tab:databaseFields}}
\end{table*}

Five complete mock catalogues for the Aquarius haloes A-E are publicly
available at the following website   
\url{http://galaxy-catalogue.dur.ac.uk:8080/StellarHalo}. Each mock is stored in a
separate table within the StellarHalo database and can be accessed via the web
interface using SQL queries. The database will contain all stars with
absolute magnitude $M_g < 7$. This cut is intended to select
bright stars most useful to studies of the distant halo, including main
sequence turn-off stars, red giant branch stars and blue horizontal branch
stars (see \ref{sec:ExampleMocks}). It excludes a very large number of faint
main sequence stars which are not accessible to surveys of the outer halo of
the Milky Way.

Table \ref{tab:databaseFields} lists the information that is provided.
Positions, velocities and apparent magnitudes\footnote{The magnitudes are based
on {\sc sdss} $u,g,r,i,z$ filters.} are given relative to both the centre of
the halo and to an observer placed in a position similar to that of the Sun. In
order to do this we first define a plane that a galactic disc would likely lie
in and then choose a position for the Sun within that plane.  We choose to
orient the disc plane perpendicular to the minor axis of the moment of inertia
of the dark matter in the inner 10 kpc, as the minor  axis is a stable orbital
axis. While the inclusion of baryons might be expected to change the shape of
the centre of the halo due to adiabatic contraction \citep{Kazantzidis:2004},
their presence is unlikely to alter the halo's orientation. We have taken the
Sun to be 8 kpc from the centre of the galaxy and moving with a velocity $(U,
V, W)_{\odot} = (11, 232, 7)$ km~s$^{-1}$ \citep{Schonrich:2010}. The observer
has been placed on the $x$-axis, though any orientation on the solar circle
would be an equally valid choice. 

In addition to the magnitudes, we provide a few intrinsic properties of stars,
such as their age, metallicity, stellar mass, surface gravity and effective
temperature. The semi-analytical galaxy formation model \galform uses
physically-motivated prescriptions to follow the evolution of the galaxies
within each halo. While \galform outputs many properties for each galaxy, it is
only the star formation and chemical history that are used to build the mocks.
These two model predictions for the baryonic components of satellites are
essential quantities required to populate the halo with stars.  The final input
are the theoretical isochrones that provide the detailed properties of the
stars. For a given age and metallicity, together with an IMF, the isochrones
provide a breakdown of these stellar properties. The type column is the
original `stage' flag in the PARSEC isochrones that denotes different phase of
stellar evolution (see Table \ref{tab:databaseFields}).

Beyond the spatial and kinematical information for individual stellar
populations, the underlying dark matter \Nbody simulations provide the history
of halo mass assembly. Each star is associated with a dark matter particle and
the subhalo to which that particle is currently bound, through the dark matter
ID and Subhalo ID. This information allows the dynamical history of the
particle to be traced back through the simulation. The satellite which brought
the star into the main halo can be identified, along with the time of accretion
of that satellite and the time at which the particle was stripped from the
satellite.  It is possible to find stellar structures and streams easily by
grouping stars according to their satellite of origin, using the TreeID field.
The `infall' redshift of the parent satellite has been provided, defined as be
the redshift when the satellite reached its maximum mass (for most objects this
correlates closely with the actual time of entry into the main FOF group and
hence with crossing the virial radius of the central halo). 

\subsection{Errors and Extinction}

For generality and simplicity, the observables presented in our
catalogue have not been convolved with sources of observational error. In order
to simulate a particular survey, the user should therefore convolve
physical quantities such as magnitude, velocity and metallicity with
appropriate error functions.  The user may also wish to impose a foreground
dust extinction model, and superpose additional source distributions to account
for contaminations of halo star selections by faint foreground stars and the
extragalactic background.

\subsection{Limitations}

While the mock catalogues we have generated are based on sophisticated
cosmological models, they are not without limitations. The biggest limitation
is that our model only generates the accreted component of the stellar halo.
It does not include processes that create {\em in situ} halo
components in hydrodynamical simulations, such as the scattering of
disc stars or star formation in gaseous streams. This means that the model is
most suitable for studying the outer regions of a galaxy $(r>20\; {\rm kpc})$.
There are various possible ways to add in additional components to create a
more complete model of the entire galaxy, and we will explore these in future
papers.  

Stars belonging to the galactic disc are not included in the mocks and the
effect of having a disc present within the halo during its evolutionary history
is also missing. Subhaloes that pass near or through a disc are likely to be
disrupted more rapidly \citep{DOnghia:2010b}. The importance of such effects is
unclear because the disc has been growing as the halo grows; even today, it
presents a small cross section to infalling satellites. This effect is
therefore most relevant to the small number of satellites that pass close to
the centre of the halo after at late times. The consequences of
neglecting enhanced disruption by the galactic disc are likely to vary among
our five simulations, according to their particular mass assembly histories and
subhalo orbital distributions. While the properties of individual streams may
change significantly, we expect that, in most cases, the overall properties of
the halo should be robust to the inclusion of a realistic disc.

As discussed in section~\ref{sec:tagging}, the C10 technique uses dark matter
particles to trace the distribution and dynamics of stars. In reality, stars
may form on phase space trajectories that are not well sampled in collisionless
simulations (for example, centrifugally supported discs). For this reason, the
spatial configuration of stars within surviving satellites may not be reliable
in detail. The importance of the initial configuration is greatly diminished
once a satellite has been disrupted and the stars phase-mixed into the halo,
but studies of chemical and kinematic gradients along coherent streams should
keep this limitation in mind.

\subsection{Example Mocks}
\label{sec:ExampleMocks}

\begin{figure}
\centerline{\includegraphics[width=1.0\linewidth]{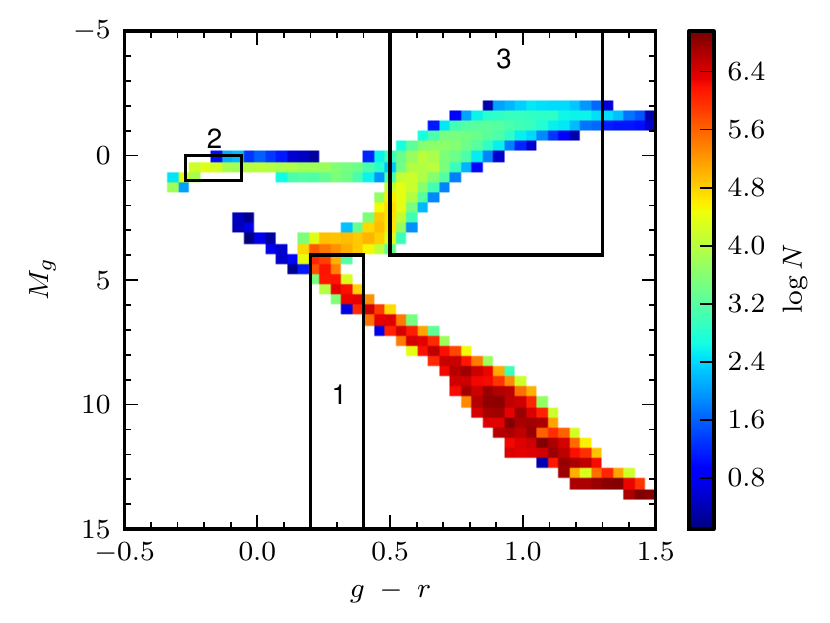}}

\caption{Colour-magnitude diagram for the overall stellar population of the
accreted stellar halo of Aq-A. Stars still bound to satellites are not
included. The colours encode the number of stars within a bin and the boxes
illustrate the \textit{approximate} regions covered in our three mock catalogue
selection functions (see text for details of the actual selection criteria,
which make use of additional photometric bands and metallicity information).
Box 1 marks our selection for MSTO stars, box 2 marks the
\citet{Bell:2010}  selection for BHB stars and box 3 the
\citet{Xue:2014} selection for K giants. Photometric
magnitudes are given in the SDSS $ugriz$ filter system.}

\label{fig:haloCMD}
\end{figure}

We now describe a set of example mock catalogues generated from the
Aquarius simulations and use them to illustrate how particular samples
of stars can be easily selected. Each sample is defined by a set of
colour and magnitude cuts that selects a particular stellar type
commonly used for surveying the stellar halo. The overall
colour-magnitude diagram for all the stars in the stellar halo of Aq-A
(with the contribution from stars still bound in satellites removed)
is shown in Fig.~\ref{fig:haloCMD}. The selections for our three
samples of halo tracers are marked by boxes.

The vast majority of stars are low mass stars that have not yet evolved off the
main sequence and account for a significant fraction of the total stellar mass.
However, as mentioned above, these are very faint. They can only be
observed in the Solar Neighborhood and are of little interest to studies
of the distant halo or its global properties. A clear red giant branch and
horizontal branch are visible; these contain the older, brighter stars which
can be observed to large galactocentric distances and which are thus are most
useful when studying the structure of the halo.

\subsubsection{MSTO}

Main sequence turn off (MSTO) stars can be unambiguously selected using the
colour cut $0.2 < g-r < 0.4$ \citep[e.g.][]{Bell:2010}, and an absolute
magnitude cut $M_g>4$, in order to only select dwarf stars and remove giants.
MSTO are relatively numerous and therefore serve as a good proxy for the
general stellar content of the stellar halo. Approximately 40\% to 50\% of the
stellar mass brighter than $M_g=7$ is found in MSTO stars and there are
81,506,853 such stars in the Aq-A halo. The volume currently accessible
using MSTO stars is a relatively small fraction of the total halo, but their
high surface density in this volume allows a great deal of information to be
gathered by wide-field photometric surveys such as SDSS
\citep[e.g.][]{Juric:2008}.

\subsubsection{BHB}

Blue horizontal branch (BHB) stars are very often used as tracers of the
distant stellar halo. They are luminous and have a very nearly constant
magnitude, $M_g \sim 0.7$, so they can be located accurately to very large
distances along the line of sight even with photometry alone. Unfortunately,the
conditions under which BHB stars form are still not well understood. This makes
their abundance in our mock catalogues highly sensitive to the assumptions made
by the theoretical isochrones we use. 

To select BHB stars in our mock catalogues we use the colour criteria proposed
by \citet{Bell:2010}: $0.98 < u-g < 1.28$, $-0.27 < g-r < -0.06$ and excluding 
the region $([u-g-0.98]/0.215)^2 + ([g-r+0.06]/0.17)^2 < 1$. With this selection, 
BHB stars make up a small fraction of the total mass in stars brighter than
$M_g=7$, ranging from 1-3\% in the five mocks. There are 239,950 BHBs in the
Aq-A halo.

The main source of stellar contamination in photometric BHB selection
is fainter blue straggler (BS) stars. The formation of blue stragglers is also
not well understood, and they are not included in the {\sc parsec} isochrones.
Hence, even halo BS (and white dwarfs) are not a source of contamination in the
mocks. This should be kept in mind when comparing to observed photometric BHB
counts, as should the absence of quasar contamination at faint magnitudes.

\subsubsection{K giants}

K giants are another stellar type frequently used to map the Milky
Way's stellar halo \citep{Morrison:1990, Morrison:2000,
  Starkenburg:2009}. They have the advantage that they are found in
predictable numbers in old populations of all metallicities. However,
unlike other standard candles, their luminosities range by two orders
of magnitude, which makes obtaining accurate distances more
challenging. K giants are one of the types that have been targeted for
spectroscopy by the Sloan Extension for Galactic Understanding and
Exploration \citep[{\sc segue},][]{Yanny:2009} and recently,
\citet{Xue:2014} have estimated distances to a sample of 6036 K 
giants using a combination of observed $g - r$ colours and spectroscopic
metallicity.

In order to isolate K giants in our stellar halo models we have used
the colour cuts $0.5 < g-r < 1.3$ and $0.5 < u-g < 3.5$ from
\citet{Xue:2014}, along with their proposed empirical polynomial
relation between $(g-r)$ and [Fe/H] to remove red horizontal branch
and red-clump (RC) giants. All stars with $(g-r) > 0.087\,[{\rm
  Fe/H}]^2 + 0.39\,{\rm [Fe/H}] + 0.96$ is excluded from the
selection. An additional cut of $M_g<4$ removes faint dwarf stars of
the same colours. Our selection results in the subset of stars
contained within box 3 of Fig.~\ref{fig:haloCMD}, with a large
fraction of the stars at the fainter end removed as being either red
horizontal branch or red-clump giants. About $6-7\%$ of the total
stellar halo mass in stars brighter than $M_g=7$ is found as K
giants and there are 956,246 K giants in the Aq-A halo.

\section{Density Profile of the Stellar Halo}

\subsection{Spherically averaged radial profiles and breaks in the stellar halo}
\label{sec:spherical_average_dp}

\begin{figure}
\centerline{\includegraphics[width=1.0\linewidth]{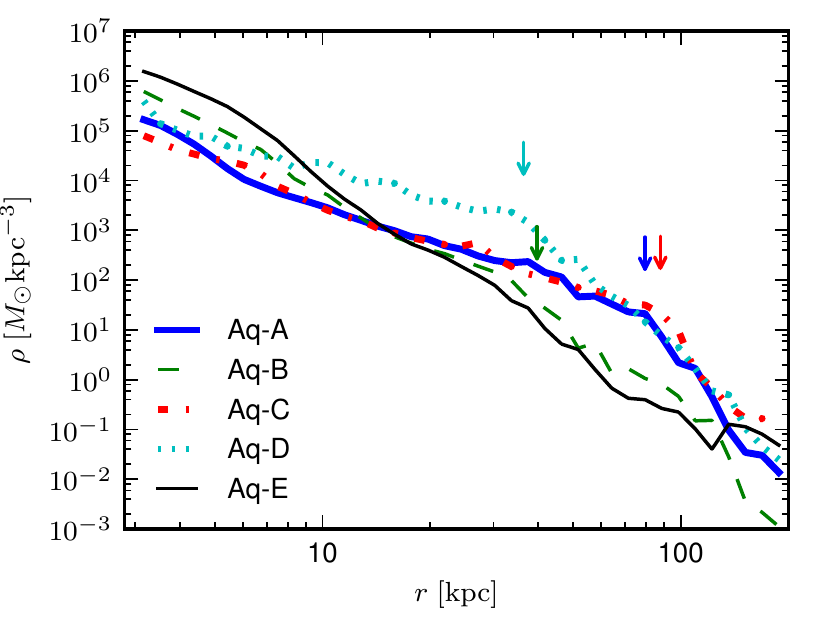}}
\caption{The spherically averaged density profiles of the five Aquarius stellar
haloes. The arrows show the location of the break in a best-fit broken power
law. Three out of the five haloes have clear breaks, a fourth has a weak break
while Aq-E is well fitted by an unbroken power law.}
\label{fig:overallDensityProfile} \end{figure}

\begin{figure}
\centerline{\includegraphics[width=1.0\linewidth]{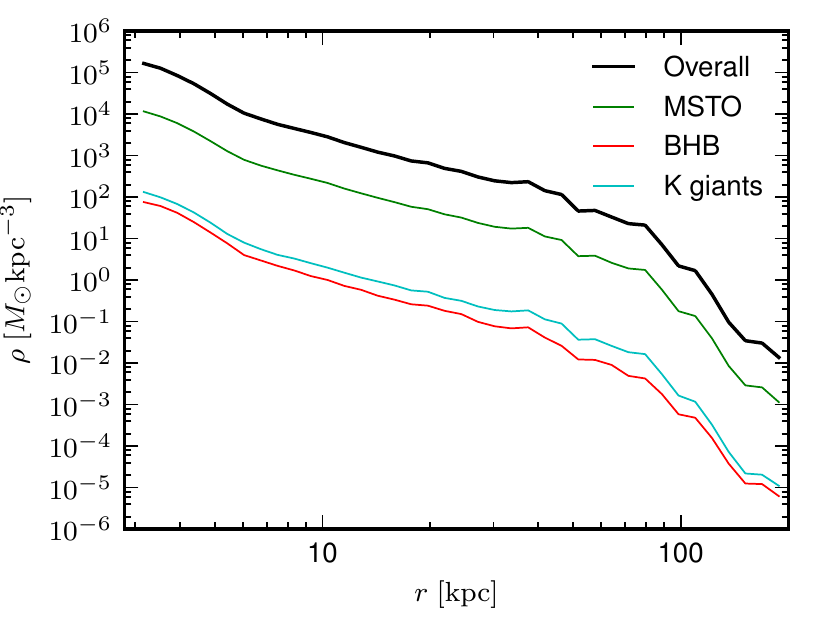} }

\caption{The spherically averaged density profile of the Aq-A stellar halo
obtained using different types of tracer stars as indicated in the legend. The
black line shows the density profile of all stars in the halo.}

\label{fig:tracerDensityProfile}
\end{figure}


One of the most basic, but still unanswered, questions in galactic structure is
the density profile and shape of the Milky Way stellar halo. While there is now
general consensus that the stellar halo is oblate and follows a broken
power-law density profile, the exact form is still ill constrained.
Evidence for a break has been found by studies using RR Lyrae variables
as standard candle tracers of the distant metal-poor halo, with measured break
radii of $\sim25$~kpc \citep{Watkins:2009}, $\sim30$~kpc, \citep{Sesar:2010}
and $\sim45$~kpc \citep{Keller:2008,Akhter:2012}.  Measurements using other
types of tracers have also suggested the existence of a break.
\citet{Deason:2011}, using a large sample of BHB and blue straggler stars from
{\sc sdss } Data Release 8 (DR8), again found the profile to follow a broken
power law, with a break radius at $\sim$27~kpc.  Similarly, from a sample of
near main-sequence turn off (MSTO) stars from the Canada-France-Hawaii
Telescope Legacy Survey ({\sc cfhtls}), \citet{Sesar:2011} measured the break
to be at $\sim$28~kpc. In contrast, no break or slope steepening was seen by
\citet{De-Propris:2010}, who analyzed a sample of 666~BHB stars in two
different directions separated by about $150^\circ$ on the sky. Instead they
found a smooth stellar distribution obeying a power-law density profile of
index $\sim$-2.5. \citet{Akhter:2012} suggested that an explanation for these
differences in the measured location of the break could be that, at larger
distances, the outer halo is not be smooth but dominated by clumps and debris.
Other studies of the power law index (or indices) of the Milky Way halo
include \cite{Chiba:2000,Miceli:2008,Carollo:2010}.

In Fig.~\ref{fig:overallDensityProfile} we show the density profiles of the
accreted stellar haloes in our five simulations, and we have excluded stars in
surviving bound subhaloes. Three of them show clear breaks, although generally
at larger distances than estimated for the Milky Way. To check whether the type
of tracer makes a difference we also show in Fig~\ref{fig:tracerDensityProfile}
the stellar density profile of the Aquarius~A halo measured from MSTO, BHB and
K giant stars. The overall density profile is that of all stellar mass in the
tags from the C10 method, while the tracer profiles are from samples taken from
the mock catalogued described in Section~\ref{sec:ExampleMocks}.  All of the
curves have the same shape, with the break occurring at the same radius, but
with differing normalizations, depending on the abundance of the stellar type.

\begin{figure}
\centerline{\includegraphics[width=1.0\linewidth]{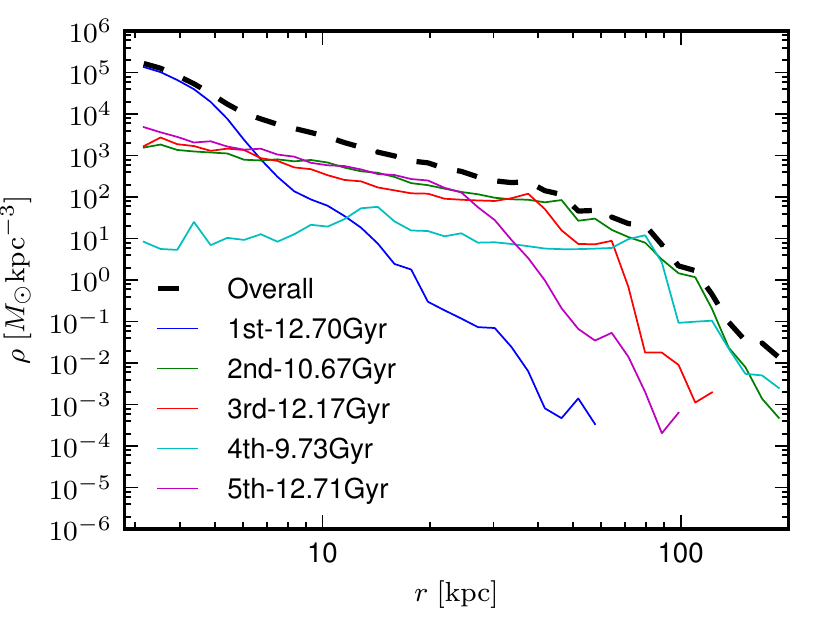}}
\caption{The density profile of the Aq-A stellar halo broken down into
the five satellites which each contribute the largest amount of mass
to the stellar halo. Numbers next to the legend show the look-back 
infall time of the five satellites.}
\label{fig:contributorsDensityProfile}
\end{figure}

It has been suggested that the presence of a break in the Milky Way's density
profile occurs at the transition between an inner region where the halo is
dominated by in-situ stars and an outer region which is primarily composed of
accreted halo stars. However, this cannot be the case for the breaks seen in
the profiles of the mock catalogues, since they do not contain any in-situ
component. Recently, \citet{Deason:2013} have proposed that the breaks
in stellar halo profiles are related to the satellite accretion history of
their galaxies and that a massive accretion event occurring 6 to 9 Gyr ago is
likely to have caused the break in Milky Way density profile. Satellites that
are just falling into the halo and are still in the process of being tidally
disrupted, may have already lost a significant fraction of their stars through
tidal stripping. Although these stripped stars are unbound and now form part of
the halo, their tidal debris is still in the form of large coherent structures
roughly tracing the orbits of their progenitor satellites. The merit of
constructing a galactocentric, spherically averaged density profile for these
debris streams and clouds is questionable, as they are far from being spherical
or dynamically relaxed. Such profiles generally appear flat in the central
regions and break to a very rapid falloff at their outer edge.

In Fig.~\ref{fig:contributorsDensityProfile} we show the density profiles of
the stars that were brought in by the five largest contributors to the Aq-A
stellar halo. Numbers beside the legends correspond to the look-back
time of infall for these satellites. Except for the very largest contributor,
the profiles of the others all have very strong breaks. A visual inspection
reveals that the largest contributor was accreted earliest at about
12.7~Gyr ago. By the end of the simulation, it has been completely disrupted
and mixed, with a correspondingly smooth profile. In contrast, the 4th largest
contributor has fallen in more recently 9.73~Gyr ago and is still in
the form of a coherent, massive stream. It has a sharp break around
$\sim80$~kpc. The other three contributors are in intermediate stages of
disruption; all show clear caustic/shell structures (visible as localized
density peaks). Notice the fifth most massive contributor shows these
structures despite having been accreted at roughly the same time as the largest
contributor, which is well mixed. The largest contributor has about 4.5 times
the mass of the fifth most massive, and hence the decay of its orbit through
dynamical friction may have been more rapid.

In Aq-A, breaks in the individual contributor profiles correspond to
these caustic structures (this is true also for the other four haloes, which we
do not plot here). This finding is consistent with the results of
\cite{Deason:2013}, that breaks are related to the apocentres of accreted
satellites. These few most massive contributors determine the overall
profile -- over certain ranges of radius, almost the entire stellar mass comes
from a single progenitor. For example, around 50~kpc the overall density
profile is almost entirely determined by the 2nd largest contributor. It is
clear that the break in the overall profile in this region is driven by stars
from this object. It can be clearly seen that the number of breaks in the
overall profile, and their sharpness, will depend strongly on the relative
distribution of stars from the few satellites that contribute significant
amounts of debris to the halo.

It can be argued that when measuring the spherically averaged stellar halo
density profile, coherent structures such as those just described should be
removed. However, while a massive stream such as the Sagittarius stream in the
Milky Way might be easy to exclude, objects in later stages of disruption are
more difficult to identify. In the simulations we are able to use the \subfind
halo finder to identify bound structures and follow the accretion histories of
stars in order to find candidate streams but this cannot be done with
observational data. In addition, even if structures have been identified, the
question is then how to decide whether they should be excluded from the halo.
For these reasons, as the overall profile is so sensitive to individual
accretion events, and thus is likely to change rapidly with time, it is not
clear how useful it is to measure the density profile of a stellar halo, other
than for the purpose of gaining information about recent large accretion
events.

\subsection{Pencil beam surveys}
\begin{figure*}
\includegraphics[width=0.8\linewidth]{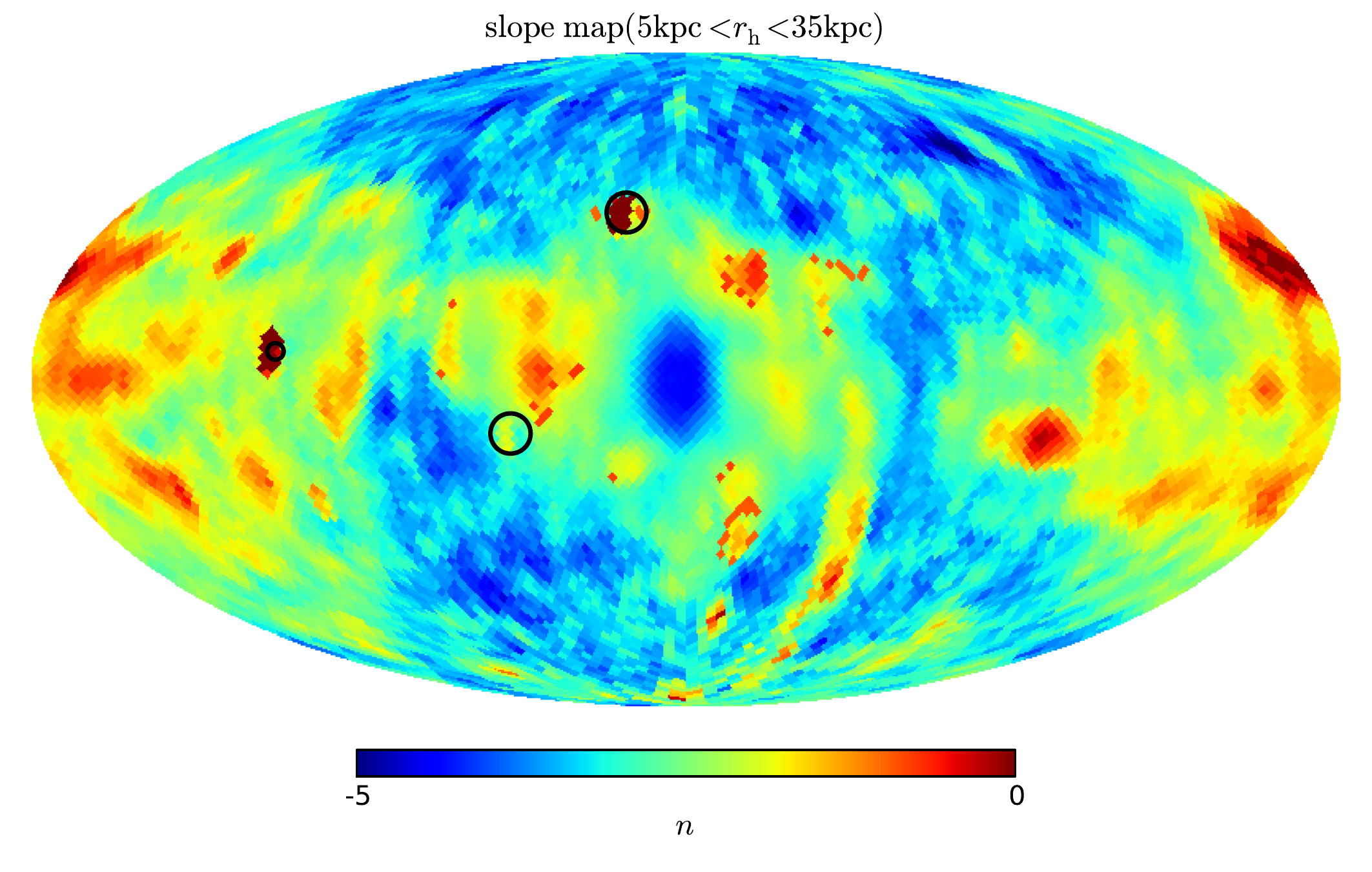}\\
\includegraphics[width=0.8\linewidth]{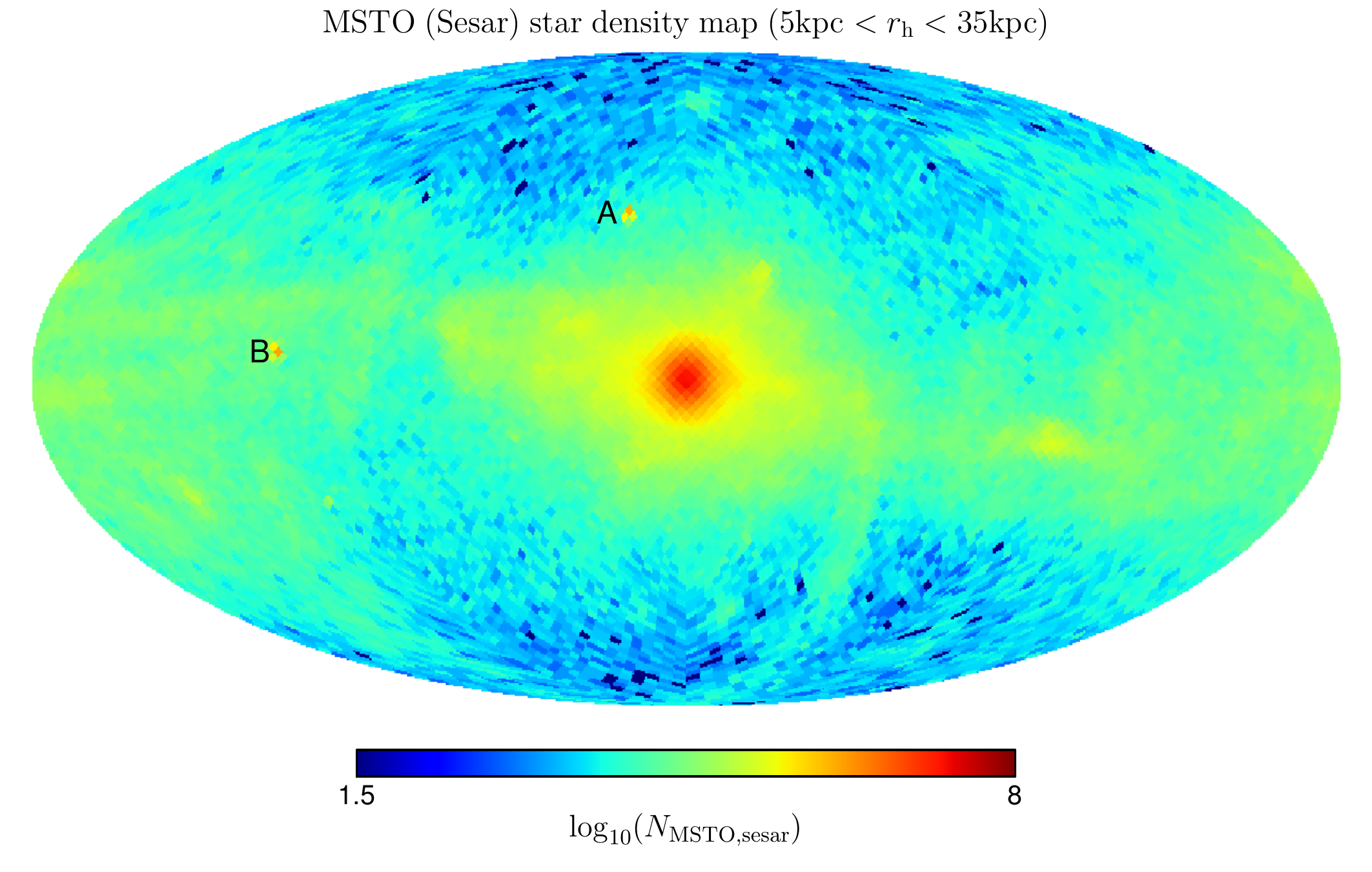}

\caption{Top: A heliocentric all-sky map in which the colour of each 8
sq. degree pixel corresponds to a value of $n$, the best-fit
(dimensionless) power law slope of the galactocentric density profile
of stars from our mock catalogues, selected according to the MSTO
colour-magnitude criteria of \citet[][see text]{Sesar:2011} in the heliocentric
distance range $5 < r_\mathrm{h} < 35$~kpc. The Galactic Centre is at the centre of
this map and our fiducial disc plane is oriented along the equator.  Unlike
previous figures, bound satellites have \textit{not} been excised in this
analysis, although there are only three in this distance range (black circles,
diameter equal to 10 times the apparent angular size of the corresponding
subhalo). Bottom: The logarithm of the projected surface
number density of stars selected and observed in the same way as the top panel.
Labels A and B mark the location of satellites discussed in the text. }

\label{fig:pencilBeamSkyA}
\end{figure*}

\begin{figure}
\centerline{\includegraphics[width=0.49\textwidth]{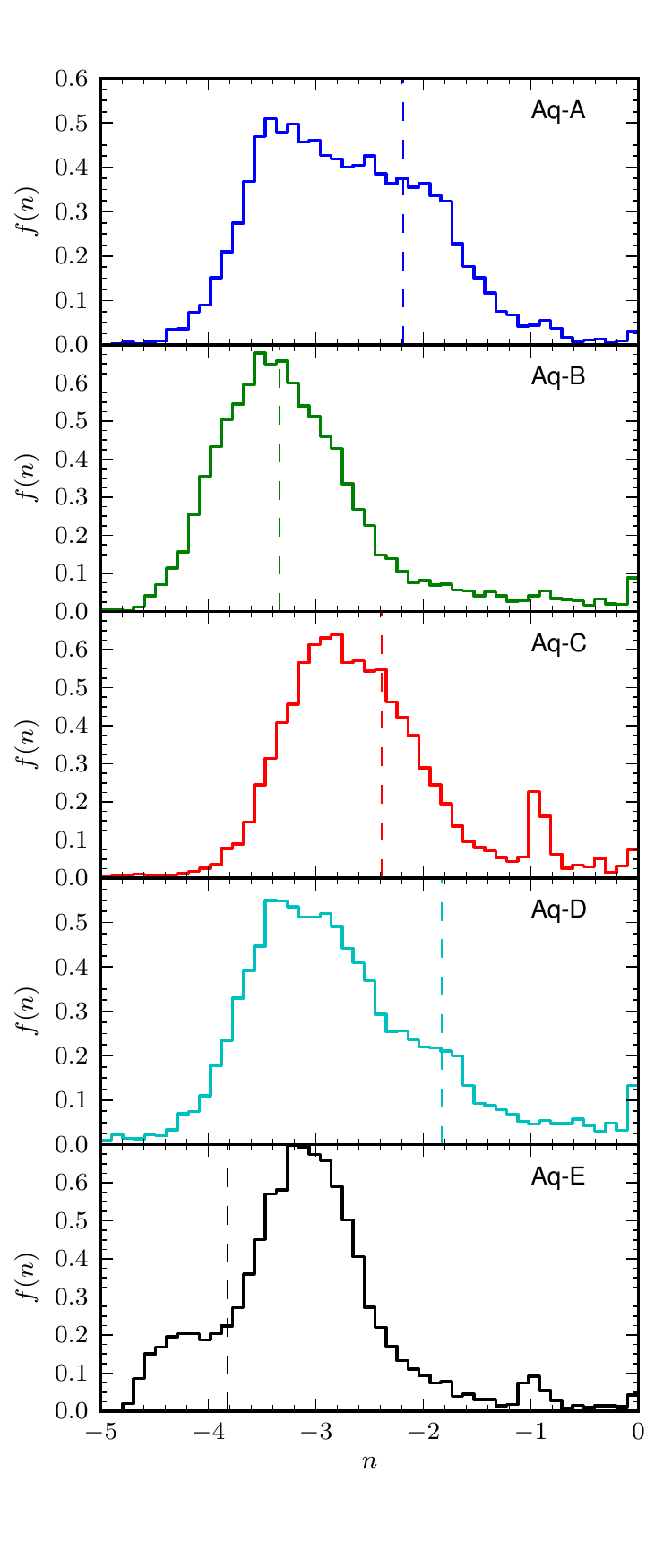}}

\caption{The distribution of volume density profile power-law slope measured
for halo stars in our mock catalogue selected using the criteria of \citet[][
see text]{Sesar:2011} in patches of area 8~sq.~degree tiling the whole sky.
Each panel shows a different mock catalogue. Bound subhaloes and other
overdensities have \textit{not} been removed; they contribute a small fraction
of the pixels in these area-weighted distributions. The dashed vertical lines
mark the overall slope from a power law fit to the (mass-weighted) spherically
averaged density profile in the same volume.} 

\label{fig:pencilBeamSlopeHist}
\end{figure}

\begin{figure*}
\includegraphics[width=0.8\linewidth]{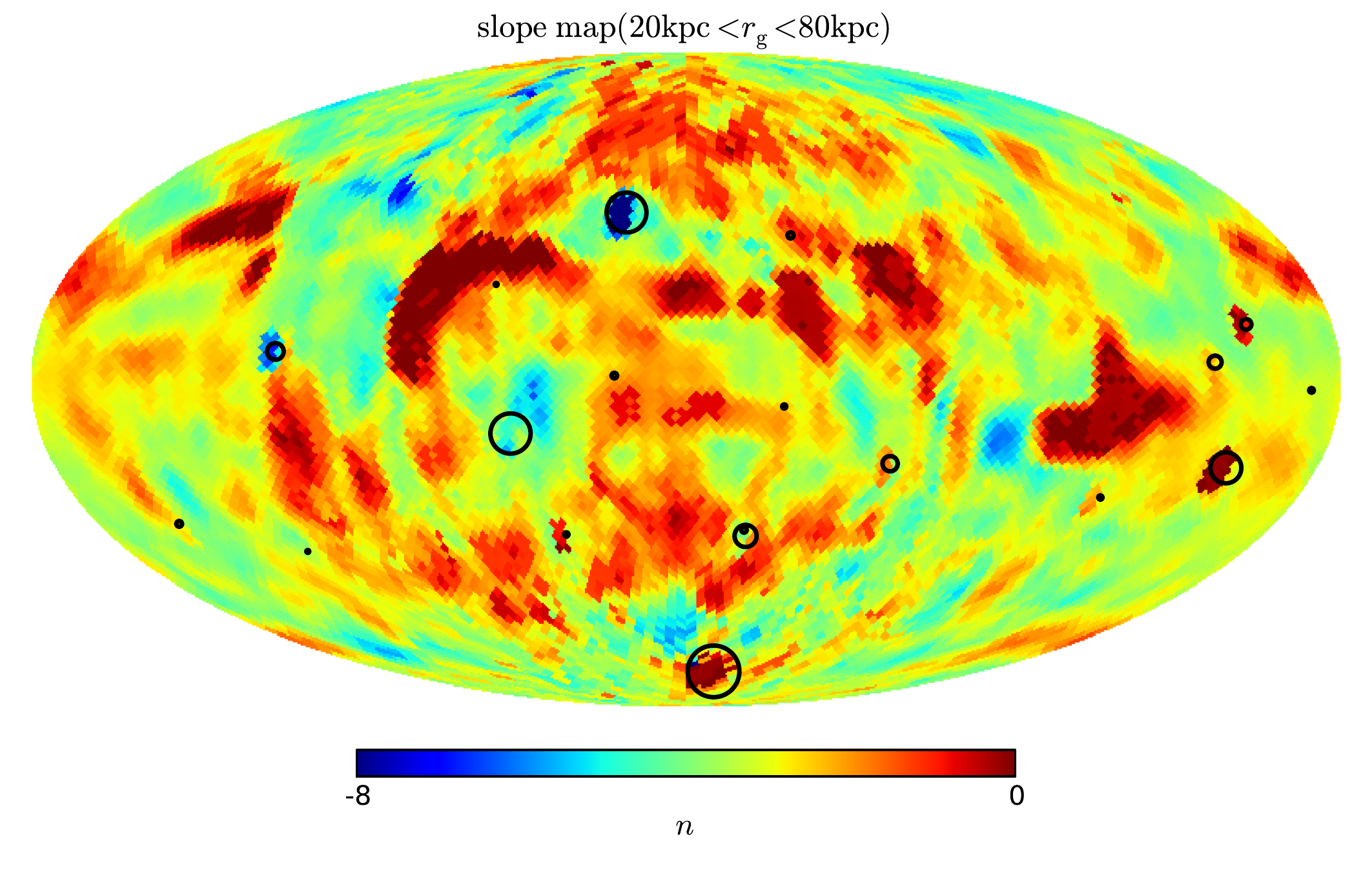}\\
\includegraphics[width=0.8\linewidth]{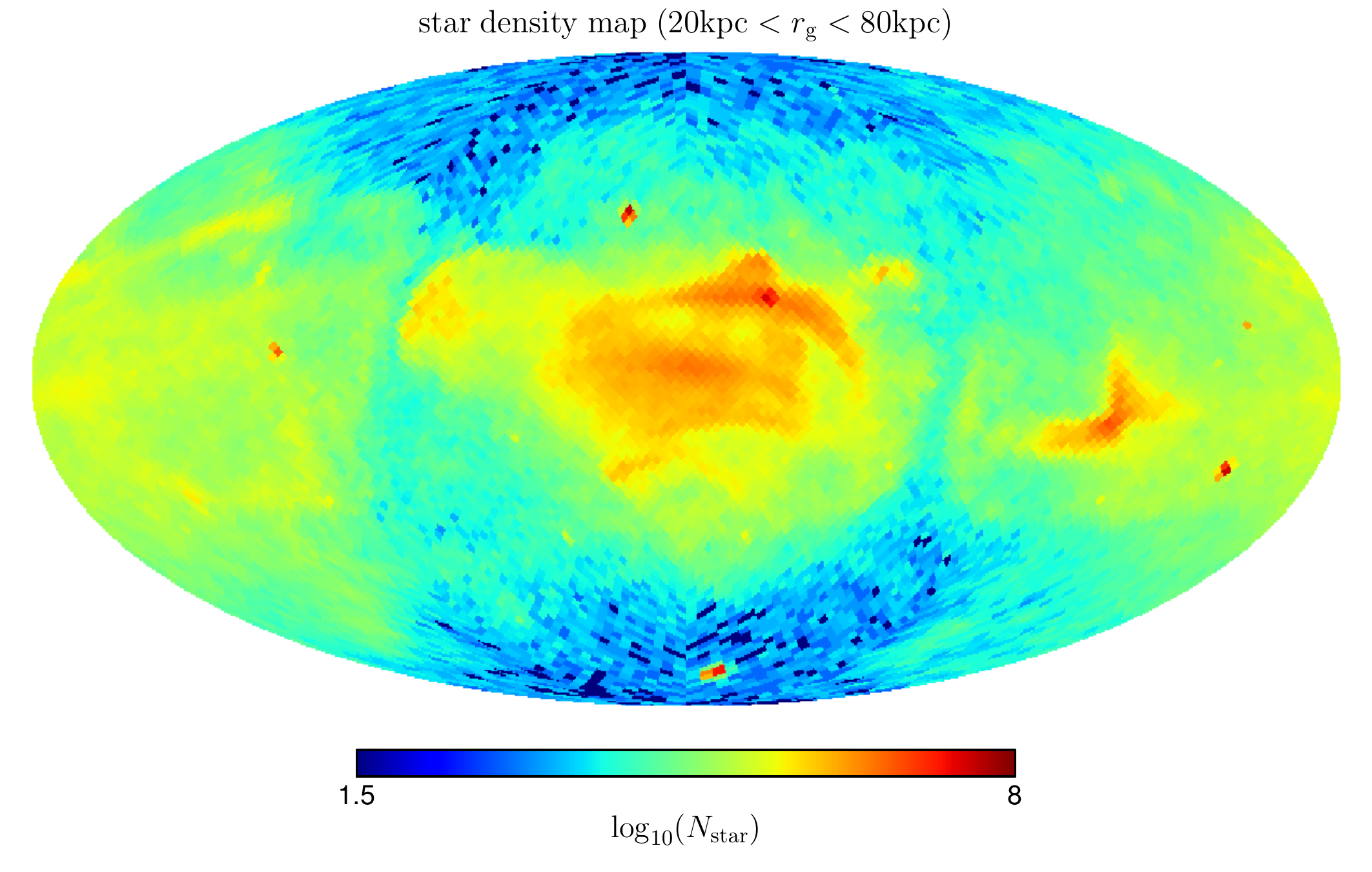}

\caption{Heliocentric sky map similar to Fig.~\ref{fig:pencilBeamSkyA}, but
showing the power law density slope (top) and surface number density (bottom)
of \textit{all} stars in our mock catalogue with galactocentric distance $20 <
r_{g} < 80$~kpc. Bound satellites and other overdensities have not been
excised. Black solid circles mark subhaloes that host satellites, with the
angular size of circle equal to 10 times the half mass radius of the subhalo as
viewed from our fiducial Solar position.}

\label{fig:pencilBeamSkyA20}
\end{figure*}

As shown in Fig.~\ref{fig:tracerDensityProfile}, the spherically averaged
density profile is not sensitive to the type of tracer used. What is much more
important is the area of the sky over which the profile is measured.  In order
to measure the profile to large radii, narrow pencil-beam surveys of
bright tracers are often used, probing deep into the stellar halo over a
limited area of sky. In order to quantify the intrinsic differences in the
slope of the profile in different directions we employ a similar observational
strategy to \citet{Sesar:2011} and measure the density the simulated halo stars
over in patches of small solid angle.  \citet{Sesar:2011} used four
`pencil beams' from the Canada-France-Hawaii Telescope Legacy Survey
({\sc cfhtls}) to select MSTO stars in order to measure the slope of the
stellar halo density profile between 5 and 35~kpc in heliocentric radius.
A direct comparison with \citet{Sesar:2011} is limited by the lack of
\textit{in situ} halo stars in our model; it may be that these dominate the
inner part of the halo.  Moreover, \citet{Sesar:2011} attempted to excise the
most obvious accreted substructures from their analysis (the Sagittarius and
Monoceros streams). This is a subjective procedure and implicitly assumes an
underlying `smooth' halo.  We do not try to excise visible streams, or even
bound satellites\footnote{Note that we remove bound satellites from the
analysis in all other sections.}. Our aim in this section is to demonstrate
how substructure in the halo affects the determination of the density profile,
rather than to compare directly with the results of \citet{Sesar:2011}.

We select stars in the same colour-apparent magnitude range as
\citet{Sesar:2011} ($0.2<g-r<0.3$,$g>17$, $17<r<22.5$, $i>17$), and measure
their galactocentric radial density profile in patches of $8\times8$
degree$^2$. Note that the \citeauthor{Sesar:2011} colour selection criteria is
slightly different from our own (simpler) `MSTO' criterion used in all other
sections of this paper (shown in box 1 of Fig.~\ref{fig:haloCMD}), and also
that the apparent magnitudes used for this selection are those for a observer
at our fiducial Solar position (see Sec.~\ref{sec:cat}). We focus on the slope
of the profile for stars in the range 5 to 35~kpc from the Sun. Although the
selection of stars is based on heliocentric distance, $r_\mathrm{h}$, the density
profiles are expressed in terms of galactocentric radius, $r_\mathrm{g}$, hence pencil
beams in different directions sample different ranges of $r_\mathrm{g}$. We tile the
entire sky with equal-area pencil beams and compute the parameters of the
best-fit power law density profile, $\rho \propto r^{n}$, in each beam. The top
panel of Fig.~\ref{fig:pencilBeamSkyA} shows a sky map of Aq-A with pixels
colour-coded by the value of $n$ along a given beam. There are large variations
in $n$ across the sky, with profiles ranging from very steep ($n\sim-5$) in
some areas to flat ($n\sim0$) in others.

The large-scale gradient in the value of $n$ with galactic latitude is
likely to result from the non-spherical overall distribution of halo stars.
This is seen clearly in the projected number density of MSTO stars (including
those bound to satellites) shown in the lower panel of
Fig.~\ref{fig:pencilBeamSkyA}. The surface density of accreted stars is high
towards the galactic centre and falls off towards the galactic poles. This
panel also shows that variations of $n$ on smaller scales (hundreds of square
degrees) correspond to localized density substructures in the halo. Some of the
most dramatic fluctuations occur around surviving galaxies that are being
tidally stripped; for example the density `hot spots' in the lower panel,
labeled `A' and `B', correspond to regions in the upper panel with $n\sim0$.
However, in general there are very few massive satellites in the range $5 <
r_\mathrm{h} < 35$~kpc; we have overplotted black circles on
Fig.~\ref{fig:pencilBeamSkyA} to highlight all those in Aq-A, with relative
diameters scaled to 10 times the angular size of their subhalos (as seen by our
fiducial Solar observer). Most small scale fluctuations in the measured density
slope, are due to inhomogeneities in the stellar halo itself rather than bound
substructures.

In order to quantify how much the slope varies over the sky, the distributions
of $n$ are shown in Fig.~\ref{fig:pencilBeamSlopeHist} for all five haloes.
These are `area weighted' distributions, in that each bin shows the
fraction of the total number of $8\deg\times8\deg$ tiles with a given slope.
In each panel, a vertical dashed line marks the overall slope of the
spherically averaged density profile in the same radial range
(effectively the mass weighted average slope). For Aq-A, B and C the
overall slope is close to either the median or \textit{mode} of the
distribution, but for Aq-D and E the overall slope is significantly different,
lying at the steep end of the distribution. This can be explained by the fact
that, as we saw earlier, a substructure or stellar stream can completely
dominate the overall profile, making it radically different from that measured
in a typical patch of the halo. It should be noted that, so far, we
have been considering the slope of the profile to no more than $35$~kpc from
the Sun. In the Milky Way, it has been suggested that an \textit{in situ}
component makes a significant contribution to the halo out to galactocentric
distances of $r_\mathrm{g}\sim20$~kpc. This component may also be much smoother than
the accreted halo originating from disrupted satellites. A smooth \textit{in
situ} component may reduce the variation in the density profile. 

To investigate the structure of the outer halo, we take all stars brighter
than $M_g=7$ in the mock catalogue with galactocentric radii in the range $ 20
< r_\mathrm{g} < 80$~kpc. This upper limit is chosen because the overall density
profiles of Fig.~\ref{fig:overallDensityProfile} do not show breaks out to
80~kpc. We again construct pencil beams to a fixed distance from the Sun and
measure the slopes of the density profiles from the galactic centre. The all
sky map of slopes is presented in the top panel of
Fig.~\ref{fig:pencilBeamSkyA20} and the surface density of stars in the lower
panel.

Compared with the top panel of Fig.~\ref{fig:pencilBeamSkyA}, the
fluctuations across the sky in Fig.~\ref{fig:pencilBeamSkyA20} are more
numerous and pronounced. The large scale variations in $n$ no longer trace
large scale variations in the surface density  of stars (bottom plot). However,
we can see many of the small scale variations in these two maps correspond.
Since stars within 20~kpc to the galactic centre are excluded, the central
overdensity in Fig.~\ref{fig:pencilBeamSkyA} disappears.

The more complex fluctuations in Fig.~\ref{fig:pencilBeamSkyA20} arise
because there are more substructures at these larger galactocentric radii, and
the local variations in the stellar halo are more pronounced. Again, we mark
all subhaloes that are more massive than $10^8M_\odot$ with black
circles. Most of these subhaloes are distant and are thus small in their
apparent size. The three most prominent local patches are associated with three
massive and extended substructures. Interestingly, the two patches that are close
to the north galactic pole and on the left hand side of the map, which can also
be identified in Fig.~\ref{fig:pencilBeamSkyA}, now give very steep slopes in
contrast to their flat slopes in Fig.~\ref{fig:pencilBeamSkyA}. This is because
the associated subhaloes are close to the Solar observer. The radial range of 5
to 35~kpc from the Sun has a very flat density distribution, beyond which the
density profile drops very quickly.

\section{Tracer Distribution}

In Section~\ref{sec:spherical_average_dp} we showed that the shape of the
spherically averaged density profiles of different stellar tracers in our mock
catalogues are similar when averaged over the whole sky. In this section we
examine the sky distribution of different stellar tracers in more detail. We
first look at the ellipsoidal shape of different populations, and then consider
the degree of variation amongst different tracers in small patches of the sky.

\subsection{The overall shape of tracers and dark matter}

\begin{figure}
\centerline{\includegraphics[width=0.49\textwidth]{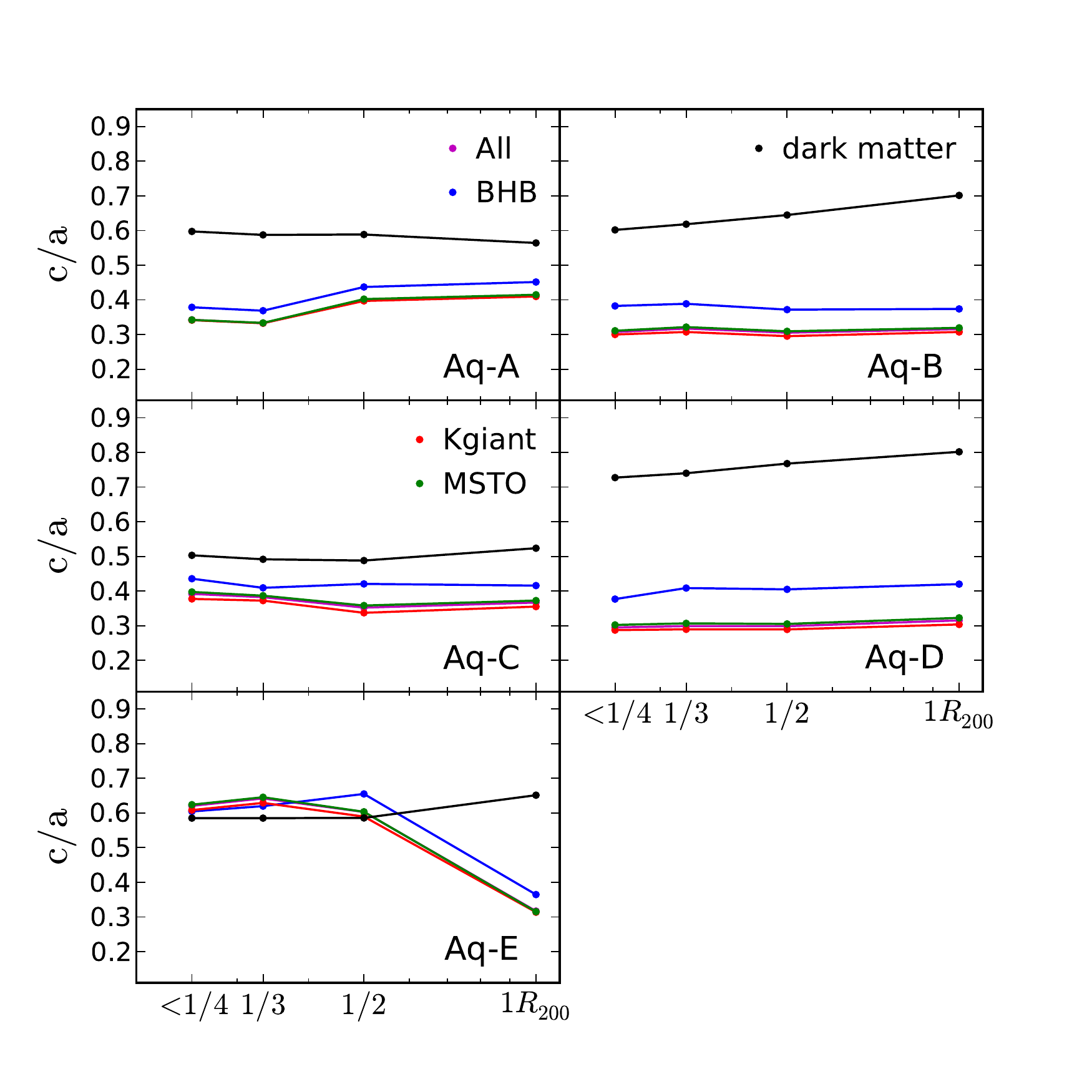}}

\caption{The overall axis ratio (minor axis versus major axis) of different
stellar tracers and the underlying dark matter, reported for four shells 
centred on the galaxy with inner radius 20~kpc in all cases and outer radii 
of $1/4$,$1/3$, $1/2$ and $1\times$ $R_{200}$ respectively.}

\label{fig:axisratio}
\end{figure}

\begin{figure}
\centerline{\includegraphics[width=0.49\textwidth]{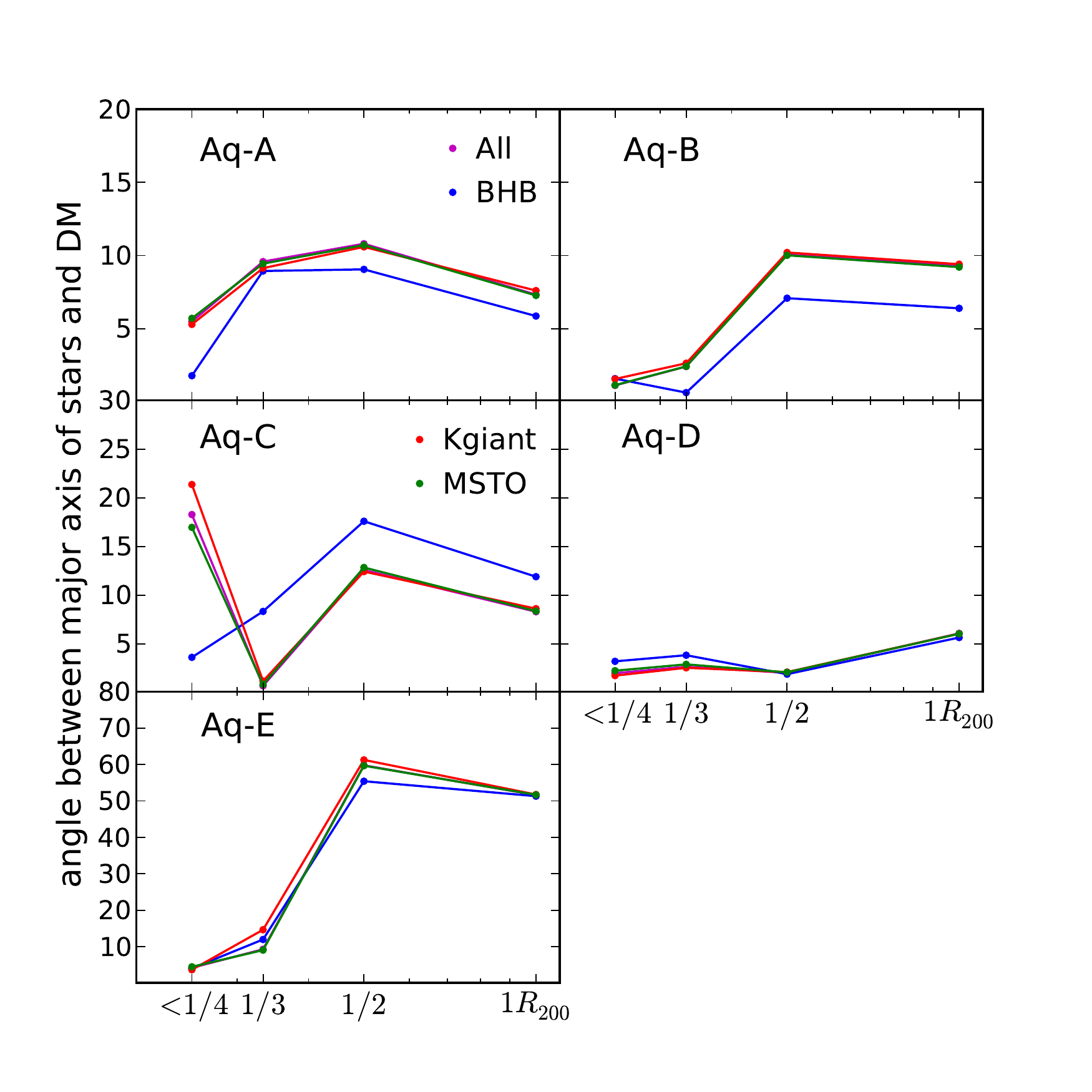}}
\caption{The angle (in units of degrees) between the major axis of stellar
tracers and the major axis of dark matter particles reported for four shells 
centred on the galaxy with inner radius 20~kpc in all cases and outer radii 
of $1/4$,$1/3$, $1/2$ and $1\times$ $R_{200}$ respectively.}

\label{fig:pointing}
\end{figure}

In this section we aim to quantify the overall ellipsoidal shape of
the stellar halo. The degree of `flattening' (or triaxiality) of the stellar
halo is of great interest, although to be meaningful, measurements based on
ellipsoidal fits to the 3d distribution of stars require that a significant
fraction of the halo be `smooth' and symmetric.

Fig.~\ref{fig:axisratio} gives the ratio between the ellipsoidal minor and
major axes, $c/a$, obtained by diagonalizing the standard inertia tensor of
different stellar populations (coloured lines, see legend) and dark matter
particles (black lines) in each simulation. We consider stars and dark matter
in four cumulative radial bins, the outer edges of which correspond to $1/4$,
$1/3$, $1/2$ and $1\times$ $R_{200}$. We exclude stars within 20~kpc of the
galactic centre and we do not include stars in any bound satellites. 

The overall shape of dark matter particles and stars are somewhat different
from each other. In the outer halo, with the exception of Aq-E, the
distribution of stars is significantly flatter than that of the dark matter.
This reflects the tendency for stars even in the outer stellar halo to be
concentrated towards the main symmetry plane of the halo, as noted by 
\citet{Cooper:2011}.
As particles at larger radii are included in the measurement, $c/a$ remains
approximately constant for stars and dark matter.  The large change in the
outermost point for Aq-E is likely to be because the stellar distribution at
large radius is dominated by only one coherent tidal stream. Interestingly, we
find that BHB stars have a slightly more isotropic distribution than the other
tracers.

Fig.~\ref{fig:pointing} illustrates how the major axes of stellar tracers
correlate with the major axes of dark matter particles\footnote{We have
checked the direction of the major axis of dark matter particles is almost
constant with radius}. For all the five haloes, we see the misalignment angle
is typically small (less than $\sim20$ degrees) particularly within
$R_{200}/4$.  Aq-E is distinguished by a sudden jump in this angle as the outer
limit of the distribution is increased, which is related to the dominant stream
mentioned above. Aq-C shows a notable misalignment between the BHB stars and
other tracers, particularly in our innermost bin. A systematic offset between
BHBs and the other tracers (such that BHBs are slightly more misaligned with
the dark matter) is also evident in Aq-A, Aq-B and Aq-C. In Aq-D, on the other hand,
all tracers of the stellar halo are very well aligned with the dark matter out
to $R_{200}$, despite the significant differences in the flattening of these
components seen in Fig.~\ref{fig:axisratio}.

\subsection{Tracer Distribution over the sky}

\label{sec:tracerdist}
\begin{figure*}
\centerline{\includegraphics[width=1.0\linewidth]{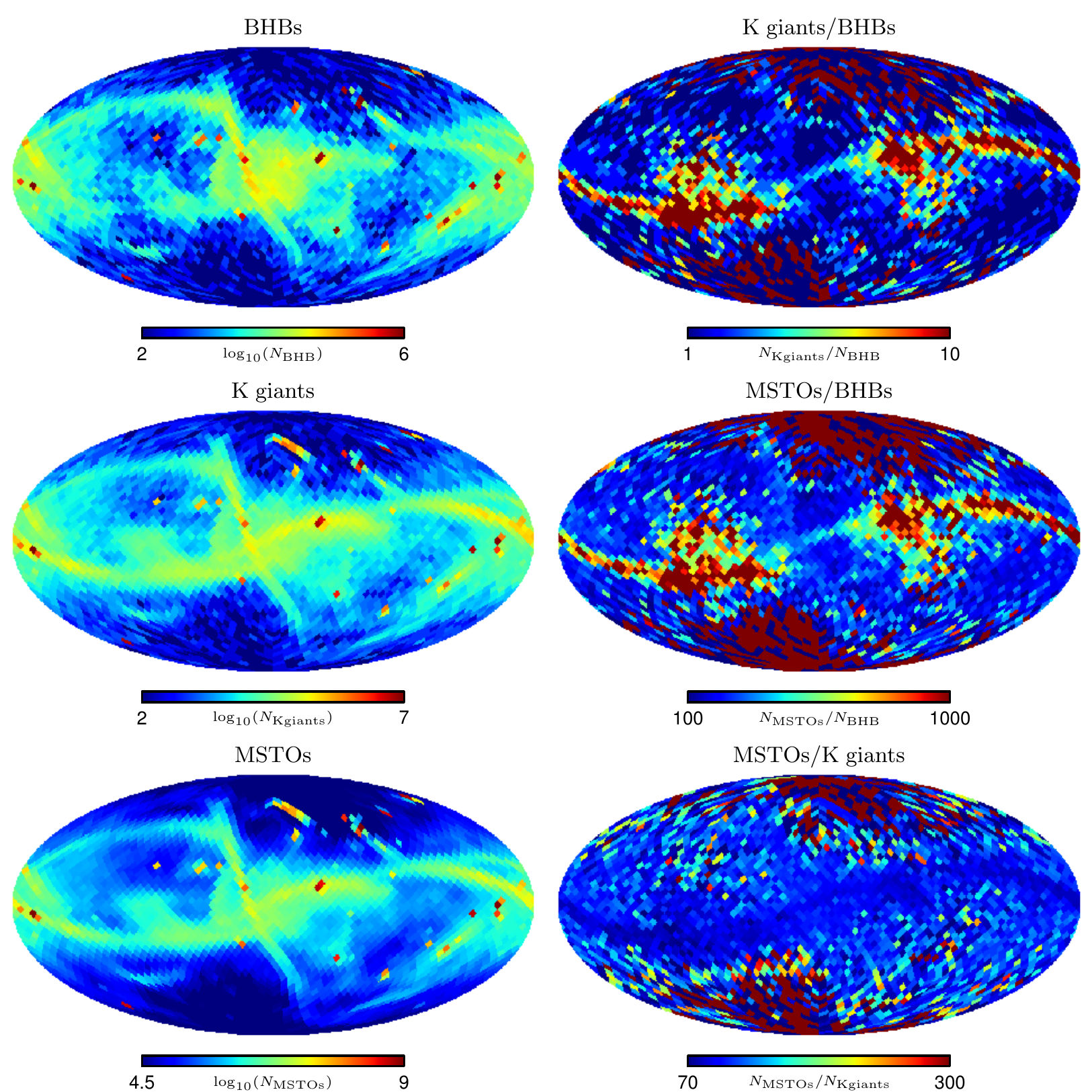}}

\caption{The distribution of different stellar tracers over the sky for the
Aq-C stellar halo as seen by our fiducial heliocentric observer. The Galactic
Centre is at the centre of the map and the disc plane is oriented along the
equator.  Stars within 20~kpc of the galactic centre have been 
excluded. {\it Left:} the projected logarithmic
number of BHBs, K giants and MSTO stars. {\it Right:} the ratio of K
giants/BHBs, MSTO/BHBs and MSTO/K giant stars.}

\label{fig:skyDistC}
\end{figure*}

\begin{figure*}
\centerline{\includegraphics[width=1.0\linewidth]{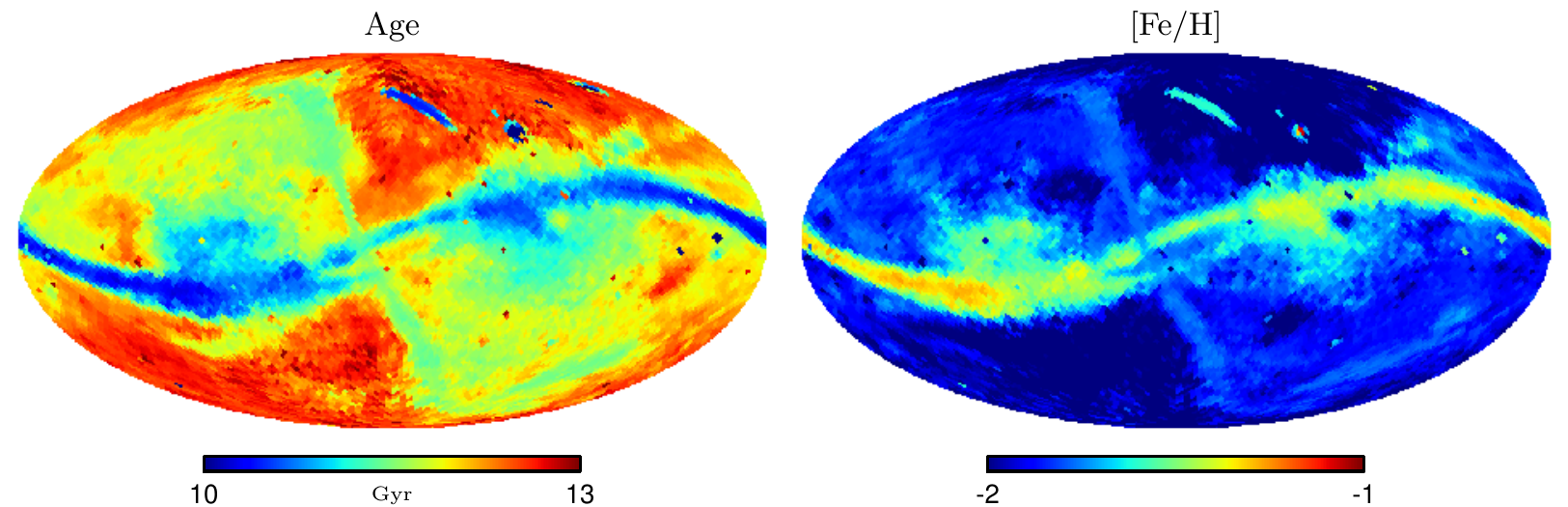}}
\caption{All-sky maps of the mean properties of the Aq-C stellar
halo when averaged along the line of sight. {\it Left map:} average
age. {\it Right map:} average metallicity map. Stars within 20~kpc of
the galactic centre have been excluded.}
\label{fig:skyPropstC}
\end{figure*}

It has been proposed that localized variations in the mix of stellar
populations may provide a method for discovering new stellar structures in the
Milky Way halo \citep[e.g.][]{Ibata:2007, McConnachie:2009}. Significant
fluctuations in the ratio of BHB to MSTO stars have been found in the stellar
halo \citep{Bell:2010}. Some of these, such as the ``low-latitude stream'' are
almost devoid of BHB stars, while other structures are rich in them.  The
Sagittarius tidal stream even shows a variation in the BHB/MSTO ratio along its
extent \citep{Bell:2010}, possibly related to a variation in mean metallicity
that has been interpreted as the result of a population gradient in the
progenitor dwarf galaxy \citep[e.g.][]{Chou:2007}. In this section we perform a
similar test with the BHB, MSTO and K giant samples from our mock catalogues
and attempt to understand the origin of the differences.

The simplest test is to search for variations by comparing sky maps of the
projected map for our different tracers as seen by a hypothetical observer
located on the solar circle. Fig.~\ref{fig:skyDistC} (left column) shows a
number of such maps for the Aq-C halo. (Aq-D and Aq-B show similar
features; to avoid repetition we focus on Aq-C only.) It contains many clearly
defined streams that are visible in all the tracers. What is most interesting,
however, is that there are two streams that do not appear in the BHB map but
appear in the two other maps. In particular, the large continuous horizontal
stream lying close to the equatorial plane that are clearly seen in the MSTO
and K giant map is partially missing in the BHB map, while the shorter stream
located near the top of the MSTO map, towards the galactic north pole, is
totally absent in the BHB map. {\textbf The absence of these streams in the BHB map
helps to explain why BHB stars show distinct features in the behaviour of axis
ratio and major axis at large radii.}

Comparing the ratio of BHBs/MSTO stars (Fig.~\ref{fig:skyDistC} right column),
the two missing streams are easily identifiable. While the large, equatorial
stream is at least partially present in the BHB map, it has a lower than
average abundance resulting in a higher MSTO/BHB ratio. In contrast, the stream
near the top of the MSTO map is almost completely absent in the BHB
map and has a huge MSTO/BHB ratio. Other streams with more standard stellar
populations have BHB to MSTO ratios similar to that of the overall halo and
thus do not show up in the ratio maps. There are also differences in the MSTO/K
giant ratio, but in the opposite sense. From the MSTO/K giant map it can be
seen that the two streams  have a higher than average abundance of K giants.
One other feature is that the BHB stars are slightly more abundant in the
galactic centre.

These differences in the relative abundance of the three types of tracer can be
explained by differences in the age and metallicity of the stellar populations
that make up the features. In Fig.~\ref{fig:skyPropstC} we show a map of the
average age (left) and the average metallicity (right) of the Aq-C stellar
halo. These have been generated from the mean age/metallicity of all stars
within each equal-area patch of the sky. It is immediately obvious that the
streams missing from the BHB map are composed of the very youngest stars in the
halo. As well as being a few gigayears younger than the mean, they are also
slightly more metal rich. In these populations BHB stars have not yet had time
to form, which explains their deficit. The actual age of these structures may
be even younger, but the average is increased by the other non-associated halo
stars along the same line-of-sight. Interestingly, there is a patch of
intermediate age stars that stands out in the lower right hand corner of the
age map, but is not visible in any of the tracer maps. This suggests that while
differences in the make up of stellar populations will help identify some
structures, not all will be found with this technique.

\section{Conclusions}

In this paper we have described a new way to construct mock catalogues
of stellar haloes based on cosmological N-body simulations. We have
adopted the particle tagging method of \citet{Cooper:2010} and
extended it to turn the output into a catalogue of individual stars
that can be directly compared to observations.

Using the Aquarius \Nbody simulations of galactic haloes as the basic
framework, we employed the semi-analytical galaxy formation model, \galform, to
calculate the evolution of the baryonic component of the Universe. This
predicts the amount and properties of the stars forming in each halo
throughout the simulation. These stellar populations were used to tag dark
matter particles in the \Nbody simulation.  Finally, these tags are turned into
a mock catalogue using theoretical isochrones to convert the fundamental
properties of the stellar populations, such as age and metallicities, into a
set of observables such as colour and magnitudes. A phase-space sampling
technique was applied to ``explode'' the massive particles into numerous
individual stars while maintaining the phase-space structure of the original
simulation.

The output of our method represents, within the limitations of the
model, a perfect dataset containing precise information about the entire
galaxy.  Actual observations usually provide an incomplete, limited view of a
galaxy.  They may be subject to selection criteria, be limited in coverage, or,
in the case of surveys of the stellar halo, be obstructed by other structures
such as the galactic disc. To make a fair comparison between simulations and
observations, it is necessary to convert the simulation data into a form that
also reproduces the limitations and biases of the observational data.
The selection criteria of an actual survey, encoded in a mask or selection
function, can be readily applied to our mocks. Observational errors,
foreground dust extinction and contaminating sources can also be
included easily. Thus, in principle, the mock catalogues can be viewed and
analyzed in a manner that mimics the observational data. In this
paper, we have used the Aquarius simulations to make mock catalogues of
galactic stellar haloes that can be used to help interpret data from upcoming
surveys such as those to be carried out by the GAIA mission.  However, there is
no reason why this technique could not be applied to other simulations of both
smaller haloes, such as dwarfs, or larger haloes, such as clusters.

Using our mock catalogues we carried out simple analyses to explore
whether current observations of the Milky Way halo provide an accurate
picture. We have found that:
\begin{itemize}

\item The accreted stellar halo is mainly built up from a few massive objects.
Those that are in early stages of disruption still maintain a coherent
structure and are locally concentrated. They dominate the spherically averaged
density profile at their corresponding radii.  Since the overall density
profile is so sensitive to these few substructures it provides limited
information about the structure of the underlying dark matter halo. Instead, it
tells us more about the recent accretion history \citep{Deason:2013}.  Three of
the five halos we have investigated show clear breaks in their spherically
averaged density profiles while two are well described by a power-law.
In our model, these breaks have nothing to do with the
distinction between accreted and in-situ components, as has often been claimed
for the Milky Way halo. Rather, they are produced by the accretion of
satellites, although not necessarily by a single one.

\item The structure of the accreted stellar halo is clumpy and thus
  varies considerably over the sky. Attempts to measure properties
  of the halo, such as its density profile, based on limited
  regions of the sky, provide biased estimates. For example, if the
  profile is modeled as a  power-law, the slopes measured in
  different pencil-beam surveys through our mock stellar haloes vary
  from very steep to flat depending on the direction of the survey. 

\item Beyond $20$~kpc, the flattening of dark matter and halo stars (as
measured by $c/a$, the major to minor axis ratio of an ellipsoidal fit) is
approximately constant out to $R_{200}$. The stellar distribution is
significantly flatter than the dark matter. The minor axis of the stellar
distribution is approximately aligned with that of the inner dark matter halo,
which we have also assumed to be the normal vector to the main symmetry plane.

\item Differences in the ages and metallicities of the satellites that build up
the stellar halo are visible as local variations in the composition of stellar
halo populations. These are reflected in differences in the abundance of
different stellar types. For example young, more metal rich objects tend to
lack BHB stars. Searching for such fluctuations in halo star surveys offers a
way to identify streams and structure in the halo.  

\end{itemize}

In this paper we have developed a method of converting massive stellar
particles into stars based on the output of the \citet{Cooper:2010} \Nbody
particle tagging method. However, with only minor modification it should be
possible to use the same conversion technique on hydrodynamical galaxy
formation simulations. By combining these two approaches, components
that are missing from our treatment, such as the galactic disc or bulge, could
be included, while other components, such as the stellar halo, could be
followed with much higher resolution than is attainable with current
hydrodynamic simulations.

The five complete mock catalogues for the Aquarius haloes A-E, are publicly
available at \url{http://galaxy-catalogue.dur.ac.uk:8080/StellarHalo}, in a database
that can by queried via SQL. The properties that are available in each
catalogue are described in the Table~\ref{tab:databaseFields}.

\section*{Acknowledgments}

We thank the anonymous referee for their careful reading of our paper
and helpful suggestions. WW is grateful for useful discussions with Jiaxin 
Han and Yanchuan Cai. This work was supported by the European Research
Council [GA 267291] COSMIWAY and Science and Technology Facilities Council
[ST/F001166/1,ST/L00075X/1]. WW is supported by the Durham International 
Junior Research Fellowship [RF040353]. APC acknowledges a Chinese Academy of Sciences
International Research Fellowship and NSFC grant no. 11350110323.  The
simulations for the Aquarius Project were carried out at the Leibniz Computing
Centre, Garching, Germany, at the Computing Centre of the Max-Planck-Society in
Garching, at the Institute for Computational Cosmology in Durham, and on the
STELLA supercomputer of the LOFAR experiment at the University of Groningen.
Sky maps were produced with the {\sc healpy} implementation of the {\sc
healpix} algorithms \citep[\url{http://healpix.jpl.nasa.gov};][]{Gorski:2005}.

\bibliographystyle{mn2e}
\bibliography{StellarHaloes}

\label{lastpage}
\end{document}